\documentclass[letterpaper,journal,twoside,romanappendices]{IEEEtran}

\usepackage{amsmath,amssymb}
\usepackage{graphicx}
\usepackage{url}
\usepackage{nicefrac}

\usepackage{tikz}
\usetikzlibrary{shapes,decorations}

\title{Adaptive Estimation-Based Safety-Critical Cruise Control of Vehicular Platoons}
\author{Vishrut Bohara and Siavash Farzan%
\thanks{Copyright \copyright\, 2024 IEEE. Personal use of this material is permitted. However, permission to use this material for any other purposes must be obtained from the IEEE by sending a request to pubs-permissions@ieee.org.}%
\thanks{Vishrut Bohara is with the Robotics Engineering Department, Worcester Polytechnic Institute, Worcester, MA 01609, USA (e-mail: vbohara@wpi.edu)}%
\thanks{Siavash Farzan is with the Electrical Engineering Department, California Polytechnic State University, San Luis Obispo, CA 93407, USA (e-mail: sfarzan@calpoly.edu)}}

\begin{document}
\maketitle
\thispagestyle{empty}
\pagestyle{empty}

\begin{abstract}
Optimal cruise control design can increase highway throughput and vehicle safety in traffic flow. In most heterogeneous platoons, the absence of vehicle-to-vehicle (V2V) communication poses challenges in maintaining system stability and ensuring a safe inter-vehicle distance. This paper presents an adaptive estimation-based control design for adaptive cruise control (ACC) that reliably estimates the states of the preceding vehicle while ensuring the autonomous vehicle operates within a safe region. Lyapunov functions and Control Barrier Functions (CBFs) are employed to design a safety-critical controller that guarantees safety despite potential estimation errors.
The proposed unified control formulation addresses limitations in the existing cruise control solutions by simultaneously ensuring safety, stability, and optimal performance.
The estimator-controller framework is implemented in scenarios with and without vehicle-to-vehicle communication, demonstrating successful performance in maintaining platoon safety and stability. 
Additionally, physics engine-based simulations reinforce both the practical viability of the proposed control framework in real-world situations and the controller's adeptness at maintaining safety amidst realistic operating conditions.
\end{abstract}


\section{Introduction}
The rapid evolution of autonomous vehicles is revolutionizing the way we commute. One crucial aspect of autonomous vehicle design is the development of effective and safe controllers. A variety of control scenarios must be addressed in the design process, including lane changes, intersections, parking, and cruise control. Among these, cruise control has attracted considerable research interest due to its wide-ranging applications in vehicle platooning, eco-driving on signalized corridors, and traffic management~\cite{wanbar18}. Two prominent types of cruise control implementation are adaptive cruise control (ACC) and connected cruise control (CCC). Designing safety-critical controllers for ACC and CCC presents significant challenges, such as handling noisy sensors and input delays while ensuring system stability~\cite{orosz16}. 
CCC enables vehicles to communicate their states, facilitating access to real-time information. However, this necessitates controllers that account for communication delays. 
Conversely, ACC systems lack communication capabilities, requiring controllers to operate independently of the preceding vehicle's actual states. This challenge necessitates the development of accurate state estimators for ACC systems.

While CCC technologies offers potential advancements, ACC remains crucial for optimal driving, particularly in scenarios with heterogeneous traffic or network limitations. CCC observer designs often depend on seamless vehicle-to-vehicle communication, leaving them vulnerable to delays or disruptions. In contrast, ACC systems operate robustly and autonomously, ensuring reliable performance in these challenging conditions. By integrating ACC as a foundational building block within CCC frameworks, a more dependable approach to autonomous driving can be constructed. This integrated strategy would leverage the strengths of both ACC and CCC, ensuring both safety and optimal performance across a wider range of real-world scenarios.

\begin{figure}[t]
    \centering
    \include{tikz-helper.tex}
    \vspace{-28pt}
    \caption{Predecessor-successor configuration in Adaptive Cruise Control.}
    \label{fig:fig1-tikz}
\end{figure}

Over the years, numerous controllers have been developed to address the challenges of ACC and CCC systems. Some research has focused on mitigating the effects of communication delays, noise, and input delays. For example,~\cite{moloro22} addresses controller design in the presence of input delay and disturbances, while~\cite{acchei18} presents a cooperative adaptive controller specifically for noisy and delay-inducing communication channels. Other work, like~\cite{wanli20}, explores state prediction for optimal control in CCC systems using linear-quadratic regulator (LQR) optimization.

Shifting focus to the ACC scenario, researchers have also developed tailored observer and controller designs.  For instance,~\cite{dirder22} proposes an acceleration estimator design for cruise control systems relying solely on distance and velocity information, employing a state feedback controller for safe distance maintenance.  Similarly,~\cite{wulu18} introduces an acceleration observer control design within a discrete-time framework. In addition,~\cite{alex19} presents a prediction model for velocity distribution, integrating it with safety constraints in a model predictive controller for energy-efficient cruise control.
Finally, research addressing the uncertainty inherent in autonomous driving includes work like~\cite{trajectory28}, which presents a trajectory prediction method for surrounding vehicles. This method considers motion uncertainty, trajectory intention uncertainty, and vehicle interactions.

Another avenue of research explores optimization and learning-based control designs for cruise control applications. For example,~\cite{linnas21} offers a comparative analysis of deep reinforcement learning (DRL) and model predictive control (MPC) approaches for ACC. Works like~\cite{roumis16} demonstrate the use of genetic algorithms to optimize a proportional integral derivative (PID) controller for cruise control. To address vehicle constraints and system adaptability,~\cite{bijan20} employs an adaptive tube-based nonlinear MPC (T-NMPC) method for robust and performance-focused economic autonomous cruise control (Eco-ACC) systems. Additionally,~\cite{wang21} showcases a DRL-based CCC controller, evaluating its performance under various traffic conditions, communication delays, reward functions, and different number of vehicles.

In cruise control applications, controllers must prioritize safety. The recent introduction of control barrier functions (CBF) offers an approach for the safety design of controllers~\cite{ametab17}.
The work in~\cite{ametab17} introduces a method that integrates control Lyapunov functions (CLFs) and CBF-based inequalities as constraints in quadratic programming (QP) problems, offering a promising avenue for maintaining safety and stability in the system. However, this approach specifically addresses a sub-problem of adaptive cruise control where the leader vehicle maintains a constant velocity, and the autonomous vehicles adhere to wheel constraints.
Building upon this foundation, the authors in~\cite{wang22} have presented a general technique to integrate observer design and estimation error with CBF for safety-critical systems.
Furthermore, CBF is employed in~\cite{alaoro23} to develop a disturbance observer-based safety-critical control framework that guarantees robustness against uncertainties. The framework is used to design a robust safety controller for CCC applications that is robust against air drag and time-varying road grades. However, as the focus remains on CCC, the literature fails to discuss the specific requirements of ACC, where integrating a state estimator with controller design is imperative for safe and stable operation.    
A method for achieving string stability with bounded speed and acceleration while ensuring safe operation is presented in~\cite{karpap23}.
Moreover, the safety evaluation and enhancement techniques for CCC using CBFs are discussed in~\cite{heoro18}. The controller design assumes that the preceding vehicle's velocity and acceleration are known through vehicle-to-vehicle (V2V) communication. However, controller design without V2V communication and string stability are not discussed.
The authors in~\cite{aCBF25} propose an adaptive CBF (aCBF) and QP-based solution to accommodate time-varying control bounds and noise in the system dynamics while also addressing the QP feasibility problem. The proposed solution mitigates the conservativeness of the CBF method, but the experiments are limited to the traffic flow with fixed preceding vehicle velocity. Similarly, the work in~\cite{SafeAC26} presents a QP design integrated with CLF for stability and CBF for safety. The design includes a Smith predictor to handle input delay, which is common in real-world scenarios. Still, the ACC problem under consideration does not include time-varying maneuvers from the preceding vehicle.
Overall, the current literature on cruise control using CBFs lacks comprehensive coverage of observer design, stability-proven closed-form controller design, and a thorough analysis of time-varying maneuvers for the ACC problem. This gap in understanding limits the broader applicability of these control methods in real-world scenarios, indicating the need for further research in these areas.

To address these shortcomings, it's important to note that although the observer design and optimization-based controllers have shown promising results in maintaining the most optimal headway, they lack stability analysis and closed-form solutions to the problem. Conversely, research dedicated to continuous-time closed-form controller design suffers from conservativeness or fails to consider different traffic flow scenarios.
The proposed controller in this manuscript addresses the limitations of the existing cruise control solutions with a comprehensive, unified algorithm. It integrates observer and safety-critical controller design, enabling the least conservative headway distance while maintaining safety and stability in diverse traffic scenarios.

In summary, the main contributions of this paper include: \\
i) An estimator design for the ACC systems to estimate both the velocity and acceleration of the preceding vehicle without communication between vehicles. \\
ii) A safety-critical controller design for ACC applications developed based on the proposed estimator. The controller design incorporates a thorough safety analysis and convergence proof using Lyapunov functions and CBFs, ensuring that the ACC system maintains safe inter-vehicle distances. \\
iii) An adaptation law for the error bounds and conservative distance of the controller design to enhance the overall system performance in maintaining the least conservative distance. \\
iv) Validation of the proposed controller by comparing the simulation results with the physics realistic AirSim environment~\cite{airsim23} experimental results, showcasing the practical utility and real-world applicability of the developed controller.

The paper is organized as follows: Section~\ref{sec:problem_description} discusses the ACC problem and system dynamics. Section~\ref{sec:controller} presents the estimator design, safety-critical controller design, stability analysis, and adaptive controller design. Section~\ref{sec:results} analyzes and comments on the results. Finally, Section~\ref{sec:conc} concludes the paper and outlines potential avenues for future work.

\section{Problem Description}\label{sec:problem_description}
This section presents a formal description of the system dynamics and the ACC problem. The cruise control problem is characterized by a platoon of $n$ autonomous vehicles and a leading vehicle. The main goal is to follow the preceding vehicle while maintaining a safe distance and platoon stability. ACC emerges as a specialized subproblem within the broader cruise control domain, where each vehicle can only perceive its velocity and the distance between itself and the vehicle ahead, with no V2V communication.

The safety objective for vehicles within the platoon is to consistently uphold a safe distance from the preceding vehicle, greater than the standstill distance $d_r$.
Simultaneously, the control objective for the cruise control problem is to maintain the optimal distance from the preceding vehicle that preserves stability while ensuring it is the least possible yet safe.
This requires the vehicle to constantly adjust its speed based on the relative velocity and position of the vehicle ahead, ensuring collision avoidance even under sudden traffic dynamics. The challenge amplifies in heterogeneous platoons where different vehicle types with varying dynamics and response rates are present. The absence of V2V communication necessitates a robust estimation-based approach that can accurately predict the state of the leading vehicle and adaptively modulate the following vehicle's control response.
Furthermore, the system must be resilient to uncertainties and estimation errors, ensuring that safety is not compromised even when the preceding vehicle's behavior is unpredictable or when environmental factors affect the sensors' accuracy.

The preliminaries for platoon dynamics and string stability are described in the following, while the controller formulation will be presented in the next section.

\subsection{Platoon Dynamics}
The dynamics of $(i-1)^{th}$ and $i^{th}$ vehicles in the platoon can be described using the predecessor-follower setup. This section discusses the predecessor-follower dynamics, while the extension to a complete platoon is addressed in the Results section. Let $d$ represent the distance between the two vehicles, $v_1$, $u_1$, and $u_j$ denote the velocity, acceleration, and jerk of the predecessor vehicle (LV), and $v$, $u$ represent the velocity and acceleration of the follower vehicle (AV) (see Fig. \ref{fig:fig1-tikz}).
The system dynamics can then be expressed as:
\begin{equation}\label{dyn:1}
    \dot d = v_1 - v ,\;\; \dot v_1 = u_1,\;\; \dot v = u,\;\; \dot u_1 = u_j
\end{equation}

To achieve maximum throughput in an ACC system, vehicles should follow their predecessors while maintaining the minimum possible distance that ensures system stability.
    The safe distance satisfying both of these conditions~\cite{swaioa94} is given by~(\ref{eq:1}), where $d_r > 0$ represents the standstill distance preset for the application and $T > 0$ denotes the time headway:
\begin{equation}\label{eq:1}
    d = d_r + Tv
\end{equation}

The standstill distance $d_r$ is the minimum spacing required between vehicles when they are completely stopped, which is preset in the design phase to prevent collisions. The time headway $T$ is the time gap between successive vehicles; a larger headway provides more reaction time, improving safety margins.

\subsection{String Stability}
Given this context, string stability~\cite{orosz16} refers to the ability of a platoon of vehicles to maintain safe inter-vehicle distances and velocities under various driving conditions without amplifying disturbances as they
propagate throughout the system.
The work in~\cite{swaioa94} indicates that knowledge of inter-vehicular distance is sufficient to ensure geometric attenuation of disturbances upstream in the vehicle platoon. Using Laplace transforms, it is demonstrated that constant headway control exhibits the desirable attribute of string stability without requiring inter-vehicle communication. Consequently, it is established that any controller maintaining a distance of $d \geq d_r + Tv$ will uphold string stability, even in the presence of non-linear disturbances in both inter-vehicular distance and vehicle velocity.

String stability is quantified using the string stability gain~($\mathcal{S}$), which measures the ratio of disturbance amplitude in the preceding vehicle to that in the following autonomous vehicle. A platoon is considered string stable if the disturbance amplitude decreases as it traverses the platoon. Consequently, platoons with a string stability gain of $\mathcal{S} < 1$ are string stable. According to this definition, a platoon system is deemed stable if it achieves string stability and the inter-vehicular distance between each predecessor-follower pair maintains a pre-established minimum safe distance $d_r + Tv$, along with a finite upper bound on this distance.

In the next section, we will employ this definition and the concept of a control barrier function to ensure that the proposed controller achieves string stability alongside the aforementioned control and safety objectives.

\section{Estimator-based Safety Critical Controller}\label{sec:controller}
In this section, we present the development of the adaptive estimation-based safety-critical controller for vehicles in an ACC setup.
Operating without relying on V2V communication, this system necessitates accurate estimation of the preceding vehicle's velocity and acceleration to optimize controller design.
This section focuses on several key aspects of the proposed system: the design of the estimator, the development of the safety-critical controller, the analysis of system stability, and the formulation of an adaptive controller design.
Fig.~\ref{fig:fig2-tikz-block} illustrates the components and their interactions in the proposed estimator-controller framework.

\begin{figure}[t]
    \centering
    \tikzset{every picture/.style={line width=0.75pt}} 

\begin{tikzpicture}[x=0.75pt,y=0.75pt,yscale=-0.75,xscale=0.75]

\draw  [color={rgb, 255:red, 74; green, 74; blue, 74 }  ,draw opacity=1 ][fill={rgb, 255:red, 0; green, 0; blue, 255 }  ,fill opacity=0.125 ] (170,85) .. controls (170,79.48) and (174.48,75) .. (180,75) -- (270,75) .. controls (275.52,75) and (280,79.48) .. (280,85) -- (280,115) .. controls (280,120.52) and (275.52,125) .. (270,125) -- (180,125) .. controls (174.48,125) and (170,120.52) .. (170,115) -- cycle ;
\draw  [color={rgb, 255:red, 74; green, 74; blue, 74 }  ,draw opacity=1 ][fill={rgb, 255:red, 0; green, 0; blue, 255 }  ,fill opacity=0.125 ] (330,85) .. controls (330,79.48) and (334.48,75) .. (340,75) -- (440,75) .. controls (445.52,75) and (450,79.48) .. (450,85) -- (450,115) .. controls (450,120.52) and (445.52,125) .. (440,125) -- (340,125) .. controls (334.48,125) and (330,120.52) .. (330,115) -- cycle ;
\draw  [color={rgb, 255:red, 74; green, 74; blue, 74 }  ,draw opacity=1 ][fill={rgb, 255:red, 155; green, 155; blue, 155 }  ,fill opacity=0.25 ] (330,155) .. controls (330,149.48) and (334.48,145) .. (340,145) -- (440,145) .. controls (445.52,145) and (450,149.48) .. (450,155) -- (450,185) .. controls (450,190.52) and (445.52,195) .. (440,195) -- (340,195) .. controls (334.48,195) and (330,190.52) .. (330,185) -- cycle ;
\draw  [color={rgb, 255:red, 74; green, 74; blue, 74 }  ,draw opacity=1 ][fill={rgb, 255:red, 155; green, 155; blue, 155 }  ,fill opacity=0.25 ] (5,15) .. controls (5,9.48) and (9.48,5) .. (15,5) -- (105,5) .. controls (110.52,5) and (115,9.48) .. (115,15) -- (115,45) .. controls (115,50.52) and (110.52,55) .. (105,55) -- (15,55) .. controls (9.48,55) and (5,50.52) .. (5,45) -- cycle ;
\draw  [color={rgb, 255:red, 74; green, 74; blue, 74 }  ,draw opacity=1 ][fill={rgb, 255:red, 155; green, 155; blue, 155 }  ,fill opacity=0.25 ][dash pattern={on 4.5pt off 4.5pt}] (10,155) .. controls (10,149.48) and (14.48,145) .. (20,145) -- (110,145) .. controls (115.52,145) and (120,149.48) .. (120,155) -- (120,185) .. controls (120,190.52) and (115.52,195) .. (110,195) -- (20,195) .. controls (14.48,195) and (10,190.52) .. (10,185) -- cycle ;
\draw  [color={rgb, 255:red, 74; green, 74; blue, 74 }  ,draw opacity=1 ][fill={rgb, 255:red, 0; green, 0; blue, 255 }  ,fill opacity=0.125 ] (170,155) .. controls (170,149.48) and (174.48,145) .. (180,145) -- (270,145) .. controls (275.52,145) and (280,149.48) .. (280,155) -- (280,185) .. controls (280,190.52) and (275.52,195) .. (270,195) -- (180,195) .. controls (174.48,195) and (170,190.52) .. (170,185) -- cycle ;
\draw   (125,65) .. controls (125,56.72) and (131.72,50) .. (140,50) .. controls (148.28,50) and (155,56.72) .. (155,65) .. controls (155,73.28) and (148.28,80) .. (140,80) .. controls (131.72,80) and (125,73.28) .. (125,65) -- cycle ;
\draw    (305,85) -- (327,85) ;
\draw [shift={(330,85)}, rotate = 180] [fill={rgb, 255:red, 0; green, 0; blue, 0 }  ][line width=0.08]  [draw opacity=0] (5.36,-2.57) -- (0,0) -- (5.36,2.57) -- cycle    ;
\draw  [dash pattern={on 4.5pt off 4.5pt}]  (120,170) -- (167,170) ;
\draw [shift={(170,170)}, rotate = 180] [fill={rgb, 255:red, 0; green, 0; blue, 0 }  ][line width=0.08]  [draw opacity=0] (5.36,-2.57) -- (0,0) -- (5.36,2.57) -- cycle    ;
\draw    (115,30) -- (305,30) ;
\draw    (305,30) -- (305,85) ;
\draw    (280,100) -- (327,100) ;
\draw [shift={(330,100)}, rotate = 180] [fill={rgb, 255:red, 0; green, 0; blue, 0 }  ][line width=0.08]  [draw opacity=0] (5.36,-2.57) -- (0,0) -- (5.36,2.57) -- cycle    ;
\draw    (305,115) -- (327,115) ;
\draw [shift={(330,115)}, rotate = 180] [fill={rgb, 255:red, 0; green, 0; blue, 0 }  ][line width=0.08]  [draw opacity=0] (5.36,-2.57) -- (0,0) -- (5.36,2.57) -- cycle    ;
\draw    (305,115) -- (305,170) ;
\draw    (280,170) -- (305,170) ;
\draw    (390,125) -- (390,142) ;
\draw [shift={(390,145)}, rotate = 270] [fill={rgb, 255:red, 0; green, 0; blue, 0 }  ][line width=0.08]  [draw opacity=0] (5.36,-2.57) -- (0,0) -- (5.36,2.57) -- cycle    ;
\draw    (140,30) -- (140,47) ;
\draw [shift={(140,50)}, rotate = 270] [fill={rgb, 255:red, 0; green, 0; blue, 0 }  ][line width=0.08]  [draw opacity=0] (5.36,-2.57) -- (0,0) -- (5.36,2.57) -- cycle    ;
\draw    (280,85) -- (295,85) ;
\draw    (295,65) -- (295,85) ;
\draw    (295,65) -- (158,65) ;
\draw [shift={(155,65)}, rotate = 360] [fill={rgb, 255:red, 0; green, 0; blue, 0 }  ][line width=0.08]  [draw opacity=0] (5.36,-2.57) -- (0,0) -- (5.36,2.57) -- cycle    ;
\draw    (140,80) -- (140,100) ;
\draw    (140,100) -- (167,100) ;
\draw [shift={(170,100)}, rotate = 180] [fill={rgb, 255:red, 0; green, 0; blue, 0 }  ][line width=0.08]  [draw opacity=0] (5.36,-2.57) -- (0,0) -- (5.36,2.57) -- cycle    ;
\draw    (140,100) -- (140,155) ;
\draw    (140,155) -- (167,155) ;
\draw [shift={(170,155)}, rotate = 180] [fill={rgb, 255:red, 0; green, 0; blue, 0 }  ][line width=0.08]  [draw opacity=0] (5.36,-2.57) -- (0,0) -- (5.36,2.57) -- cycle    ;
\draw    (120,30) -- (120,115) ;
\draw    (145,115) -- (167,115) ;
\draw [shift={(170,115)}, rotate = 180] [fill={rgb, 255:red, 0; green, 0; blue, 0 }  ][line width=0.08]  [draw opacity=0] (5.36,-2.57) -- (0,0) -- (5.36,2.57) -- cycle    ;
\draw    (120,115) -- (135,115) ;
\draw  [draw opacity=0] (135,115) .. controls (135,115) and (135,115) .. (135,115) .. controls (135,112.51) and (137.24,110.5) .. (140,110.5) .. controls (142.76,110.5) and (145,112.51) .. (145,115) -- (140,115) -- cycle ; \draw   (135,115) .. controls (135,115) and (135,115) .. (135,115) .. controls (135,112.51) and (137.24,110.5) .. (140,110.5) .. controls (142.76,110.5) and (145,112.51) .. (145,115) ;  

\draw (165,82) node [anchor=north west][inner sep=0.75pt]  [font=\small] [align=left] {\begin{minipage}[lt]{65pt}\setlength\topsep{0pt}
\begin{center}
State estimator\\\textit{{\footnotesize Eq. (3)}}
\end{center}

\end{minipage}};
\draw (3,11) node [anchor=north west][inner sep=0.75pt]  [font=\small] [align=left] {\begin{minipage}[lt]{63pt}\setlength\topsep{0pt}
\begin{center}
Sensor\\measurements
\end{center}

\end{minipage}};
\draw (167,145.5) node [anchor=north west][inner sep=0.75pt]  [font=\small] [align=left] {\begin{minipage}[lt]{62pt}\setlength\topsep{0pt}
\begin{center}
Adaptive error
\vskip -1pt
dyn. integrator
\vskip -1pt\textit{{\footnotesize Eq. (28)}}
\end{center}

\end{minipage}};
\draw (4,151) node [anchor=north west][inner sep=0.75pt]  [font=\small] [align=left] {\begin{minipage}[lt]{66.5pt}\setlength\topsep{0pt}
\begin{center}
Communication\\readings
\end{center}

\end{minipage}};
\draw (340,75) node [anchor=north west][inner sep=0.75pt]  [font=\small] [align=left] {\begin{minipage}[lt]{57.82pt}\setlength\topsep{0pt}
\begin{center}
Safety-critical
\vskip -1pt
controller
\vskip -1pt
\textit{{\footnotesize Eqs. (30) \& (31)}}
\end{center}

\end{minipage}};
\draw (320,152) node [anchor=north west][inner sep=0.75pt]  [font=\small] [align=left] {\begin{minipage}[lt]{77.22pt}\setlength\topsep{0pt}
\begin{center}
Low-level\\actuator control
\end{center}

\end{minipage}};
\draw (135,172) node [anchor=north west][inner sep=0.75pt]  [font=\small]  {$v_{1}$};
\draw (212,10) node [anchor=north west][inner sep=0.75pt]  [font=\small]  {$d,v$};
\draw (144,32) node [anchor=north west][inner sep=0.75pt]  [font=\small]  {$d$};
\draw (144,78) node [anchor=north west][inner sep=0.75pt]  [font=\small]  {$\tilde{d}$};
\draw (220,43.4) node [anchor=north west][inner sep=0.75pt]  [font=\small]  {$\hat{d}$};
\draw (372,129) node [anchor=north west][inner sep=0.75pt]  [font=\small]  {$u$};
\draw (287,155) node [anchor=north west][inner sep=0.75pt]  [font=\small]  {$\epsilon $};
\draw (285,101) node [anchor=north west][inner sep=0.75pt]  [font=\small]  {$\hat{v}_{1}$};
\draw (132.5,49) node [anchor=north west][inner sep=0.75pt]  [font=\scriptsize]  {$-$};
\draw (140,58.4) node [anchor=north west][inner sep=0.75pt]  [font=\scriptsize]  {$+$};
\draw (122,35) node [anchor=north west][inner sep=0.75pt]  [font=\small]  {$v$};

\end{tikzpicture}
    \vspace{-20pt}
    \caption{Block diagram of the proposed estimation-based control design illustrating various components in the framework.
    The components shown as dashed are only applicable in scenarios involving V2V communication.}
    \label{fig:fig2-tikz-block}
\end{figure}

\subsection{Estimator Design}
Let $\hat{d},\hat{v}$, and $\hat{u}$ represent the estimate of the state $d,v$, and $u$ respectively, and $\tilde{d}$, $\tilde{v}$, and $\tilde{u}$ represent the error in the estimate, i.e., $\tilde{d} = \hat{d} - d$, $\tilde{v} = \hat{v} - v$, and $\tilde{u} = \hat{u} - u$, and $g_1$, $g_2$, $g_3$ be the constant estimator gains. Using this notation and the distance between the vehicles $d$ sensed by the AV, the state estimator dynamics is proposed as follows:
\begin{equation}\label{dyn:2}
\begin{aligned}
    \dot{\hat d} &= \hat v_1 - v  + g_1 \tilde{d} \\
    \dot{\hat{v}}_1 &= g_2 \tilde{d} + \hat{u}_1 \\
    \dot{\hat{u}}_1 &= g_3 \tilde{d}
\end{aligned}
\end{equation}

The estimator dynamics in (\ref{dyn:2}) mirror the system dynamics, augmented with the measurement error scaled by the estimator gains. This integration of measurement error steers the estimator's prediction toward the actual measurements while maintaining resilience against measurement noise, ensuring a robust and accurate estimation.

Utilizing the estimator dynamics (\ref{dyn:2}) and system dynamics (\ref{dyn:1}), the error dynamics for the estimation error $\tilde{e} = [\tilde{d}, \, \tilde{v}_1, \, \tilde{u}_1]^T$ can be formulated as: 
\begin{equation}\label{eq:2}
\dot{\tilde{e}} =
 \begin{bmatrix}
     g_1 & 1 & 0 \\ 
     g_2 & 0 & 1 \\
     g_3 & 0 & 0 
 \end{bmatrix}
  \tilde{e} + 
  \begin{bmatrix}
     0 \\ 
     0 \\
     -1
 \end{bmatrix}
 u_j =  A \tilde{e} + B u_j
\end{equation}
Upper bounds on the error estimates of velocity and acceleration are introduced as $E_v \geq \tilde{v}_1$ and $E_u \geq \tilde{u}_1$, where $E_v,\, E_u>0$.  It is important to note that $E_v$ and $E_u$ represent upper bounds rather than absolute bounds. The choice of $E_v$ and $E_u$ affects the stability and equilibrium state of the system. The relation between these upper bounds and vehicle constraints is calculated in Section~\ref{sec:SA}, which aids in selecting suitable values.

\textit{Assumption 1.} To maintain the error estimates within bounds, the estimated error should be close to zero at $t=0$. To achieve this, $\tilde{v}$ and $\tilde{u}$ are assumed to be zero at $t=0$. This can be achieved by initiating communication with the predecessor vehicle (in the case of CCC) or starting from rest, i.e., $v_1 = u_1 = 0$. This is a reasonable assumption for the ACC applications due to the fact that platoons generally originate from a designated location or commence their operation from a state of rest within a traffic environment.

\subsection{Safety-Critical Controller Design}\label{sec1}
The optimal controller design for the ACC problem is a safety-critical controller that ensures string stability while minimizing conservativeness. According to~\cite{swaioa94}, any controller capable of maintaining the inter-vehicle distance greater than or equal to $d_r + Tv$ as in (\ref{eq:1}), with a reasonable choice of time headway $T>0$, ensures string stability. To design such a controller, we utilize the concept of control barrier functions~\cite{ametab17}.

For the given system, the function: 
\begin{equation}\label{eq:h-eq}
    h(d,v) = d-d_r-Tv
\end{equation}
characterizes the conservative behavior of the controller. The convergence and steady-state behavior of function $h$ can indicate the effectiveness of the control law. For an ideal controller, $h(d,v) = 0, \,\forall t \geq 0$. Using this function, the safe set $\mathcal{C}_1\subset \mathbb{R}^2$ is defined as: 
\begin{equation}\label{eq:3}
    \mathcal{C}_1= \{ (d,v) \in \mathbb{R}^2 : h(d,v) \geq 0\}
\end{equation}

\textit{Definition 1:} Given a safety set $ \mathcal{C}_1\in \mathbb{R}^2$ defined by (\ref{eq:3}), the continuously differentiable function $h: \mathbb{R}^2 \xrightarrow{} \mathbb{R}$ in (\ref{eq:h-eq}) is considered a \textit{safety function} if there exists an extended class $\mathcal{K}$ function~\cite{khalil17} denoted by $\alpha$ such that:
\begin{equation}
    \dot h = v_1 - v - Tu \geq -\alpha (h)
\end{equation}
Hence, any $u$ satisfying equation (\ref{eq:8}) ensures safety:
\begin{equation}\label{eq:8}
    \hat v_1 - \tilde{v}_1 - v + \alpha (h) \geq Tu
\end{equation}
With $\kappa = \frac{1}{T}$ and $E_v \geq \tilde{v}_1$, the proposed controller design is defined as:
\begin{equation}\label{eq:cl1}
    u =  \kappa \hat v_1 - \kappa E_v - \kappa v + \kappa \alpha (h)
\end{equation}
where $\tilde{v}_1$ denotes the error in estimating the velocity of the preceding vehicle. Substituting $\tilde{v}_1$ with its upper bound $E_v$ enhances the controller's robustness against uncertainties in system dynamics and measurement noise at the cost of increased conservativeness. This conservative nature is further mitigated by incorporating adaptive dynamics into the upper bound $E_v$, as will be discussed in Section  \ref{sec:aed}. 

The controller proposed in (\ref{eq:cl1}) is a continuous and non-singular function of the state estimates and sensor measurements of the follower vehicle. 
It is designed in accordance with platoon stability conditions, thereby ensuring inherent stability. To that end, the stability response of the controller will be further analyzed under various scenarios in Section~\ref{sec:SA} to establish the conditions on $E_v$ and $\alpha(h)$ that guarantee stability.

Prior to conducting the stability analysis of the system, the estimator dynamics for the safety function $h(d,v)$ in~(\ref{eq:h-eq}) are derived as follows:
\begin{align}
    \hat h &= h(\hat d, v) = \hat d - d_r - Tv = \tilde{d} + h(\hat d, v) \label{eq:hat1} \\
    \dot{\hat h} &= \hat{v}_1 - v + g_1 \tilde{d} - Tu = E_v  + g_1 \tilde{d} - \alpha (\hat h - \tilde{d}) \label{eq:hat2}
\end{align}

\subsection{Stability Analysis}\label{sec:SA}
Considering the error dynamics as described by (\ref{eq:2}), along with the estimates of the safety function from (\ref{eq:hat1}) and (\ref{eq:hat2}), the Lyapunov function candidates are investigated to analyze the stability of the ACC system in two possible cases.
The analysis of Case 1 demonstrates that the controller design ensures stability when the preceding vehicle's acceleration is constant. In Case 2, the principles from Case 1 are applied to show that the controller exhibits similar behavior in scenarios involving non-zero jerk, albeit with a shifted equilibrium point. The stability analysis for Case 1 is outlined as follows.

\noindent\textit{Case 1: $u_j = 0$ (Constant preceding vehicle acceleration)}

\noindent Consider the Lyapunov function candidate: 
\begin{equation}\label{eq:V1}
    V_1(\tilde{e},\hat d, v) = \frac{1}{2}(\hat h - d_c)^2 + \tilde{e}^T P \tilde{e}
\end{equation}
where $d_c$ is the conservative distance constant and ${P = P^T>0}$ is the solution of the continuous algebraic Riccati equation $A^T P + P A + Q = 0$~\cite{peter18} with $A$ given by (\ref{eq:2}), and ${Q = Q^T > 0}$.

\textit{Lemma 1:} For a given Hurwitz matrix $A$ and a positive definite matrix $Q$, there exists a unique positive definite matrix $P$ such that $A^T P + P A + Q = 0$.

Lemma 1 implies that $V_1$ in (\ref{eq:V1}) is a valid Lyapunov function candidate if the state matrix $A$ is Hurwitz. This requirement imposes a constraint on the estimator design, necessitating that the estimator gains $g_1,\,g_2,\,g_3 < 0$ be negative, as elaborated in Appendix~\ref{sec:A1}.

\textit{Theorem 1:} Given the Lyapunov function in (\ref{eq:V1}), the control law in (\ref{eq:cl1}), and the safety function estimate dynamics in (\ref{eq:hat2}), the error estimate $\tilde{e}$ in (\ref{eq:2}) converges to zero and the conservative distance $h$ converges to $d_c$, for the specific choice of class $\mathcal{K}$ function $\alpha$ and distance constant $d_c$.

\textit{Proof:} Computing the time derivative of $V_1$, the following is obtained:
\begin{align}
    \dot{V}_1(\tilde{e},\hat d, v) &= (\hat h - d_c) \dot{\hat h}+ \dot {\tilde{e}}^T P \tilde{e} + \tilde{e}^T P \dot{\tilde{e}} \nonumber \\
    &= (\hat h - d_c)\dot{\hat{h}} + \tilde{e}^T ( A^T P + P A)\tilde{e} \nonumber \\
    &= (\hat h - d_c)(E_v  + g_1 \tilde{d} - \alpha (\hat h - \tilde{d} )) - \tilde{e}^T Q \tilde{e}
\end{align}
By choosing the class $\mathcal{K}$ function as $\alpha (x) = -g_1x$, we have:
\begin{equation}\label{eq:12}
    \dot{V}_1 = (\hat h - d_c)(E_v  + g_1 d_c) - (\hat h - d_c )\alpha (\hat h - d_c ) - \tilde{e}^T Q \tilde{e}
\end{equation}
From (\ref{eq:12}), $\dot{V}_1$ becomes negative definite by selecting $d_c$ as:
\begin{equation}\label{eq:dc1}
    d_c = -\frac{E_v}{g_1}
\end{equation}
which results in:
\begin{equation}
    \dot{V}_1(\tilde{e},\hat d, v) = - (\hat h - d_c )\alpha (\hat h - d_c ) - \tilde{e}^T Q \tilde{e} < 0
\end{equation}
Since $\dot{V}_1$ is negative definite, it proves that the estimated error $\tilde{e}$ and conservative distance $h$ asymptotically converge to zero and $d_c$, respectively, where the steady state value of $h$ is defined as the \emph{conservative bias} of the system.

\noindent\textit{Case 2: $u_j \neq 0$ (non-zero preceding vehicle jerk)}

\noindent For this case, as the preceding vehicle jerk is non-zero, the equilibrium point for the estimation error deviates from the origin. The shifted equilibrium point is obtained by setting $\dot{\tilde{e}} = 0$ in~(\ref{eq:2}), resulting in:
\begin{align}\label{eq:res1}
    \tilde{e}^* &=
    \begin{bmatrix}
        1 \\ 
        -g_1 \\
        -g_2
    \end{bmatrix} \frac{u_j}{g_3} 
\end{align}

It is important to note that $u_j$ is used only to present the convergence guarantee of the safety-critical controller and estimator design. Similar to $d_c$, $u_j$ is not part of either the controller or the estimator dynamics. The derivation below shows that the system remains stable for any non-zero value of $u_j$. Therefore, the ACC system operating under the designed controller remains stable without any information on the actual value of the preceding vehicle's jerk $u_j$.

Consider the Lyapunov function candidate:
\begin{equation}\label{eq:V2}
    V_2(\tilde{e},\hat d, v) = \frac{1}{2}(\hat h - d_c)^2 + \xi^T P \xi
\end{equation}
with the positive-definite matrix $P$ in (\ref{eq:V1}) and $\xi = \tilde{e} - \tilde{e}^*$.
By following the same calculations as Case 1 for $V_2$, the following is obtained:
\begin{align}
    \dot{V}_2(\tilde{e},\hat d, v) &= (\hat h - d_c) \dot{\hat h}+ \dot {\xi}^T P \xi + \xi^T P \dot{\xi} \nonumber \\
    &= (\hat h - d_c)\dot{\hat{h}} + \xi^T ( A^T P + P A)\xi \nonumber \\
    &= (\hat h - d_c)(E_v  + g_1 \tilde{d} - \alpha (\hat h - \tilde{d} )) - \xi^T Q \xi
\end{align}
By choosing the same $\alpha(x)$ and $d_c$ as Case 1, we obtain:
\begin{equation}
    \dot{V}_2(\tilde{e},\hat d, v) = - (\hat h - d_c )\alpha (\hat h - d_c ) - \xi^T Q \xi < 0
\end{equation}
Since $\dot{V}_2$ is negative definite, it proves that $\xi$ converges to zero and $\hat h$ converges to $d_c$. Substituting these results into (\ref{eq:hat1}), the equilibrium value for $h$ is obtained as:
\begin{equation}\label{eq:11}
    h^* = \hat{h}^* - \tilde{d}^* = \frac{ - E_v}{g_1} - \frac{u_j}{g_3} 
\end{equation}
Substituting the safety bounds $h^* \geq 0$, $E_v \geq \tilde{v}$, and $E_u \geq \tilde{u}$ into (\ref{eq:11}) while having $\xi \xrightarrow{} 0$ gives:
\begin{equation}
   E_v \geq -\frac{g_1 u_j}{g_3} \;,\; E_u \geq - \frac{g_2 u_j}{g_3}
\end{equation}
Based on these bounds, a relationship between minimum $u_j$ $(U_{min})$, $E_v$, and $E_u$ is established:
\begin{equation}\label{eq:ub}
     E_v \geq -\frac{g_1U_{min}}{g_3} \;,\; E_u \geq -\frac{g_2U_{min}}{g_3}
\end{equation}
Although, for ACC applications, $u_j$ is not known, $U_{min}$ is the lower bound on the jerk, which, in most cases, is a known vehicle property. If $U_{min}$ is unknown, any value guaranteed to be the lower bound can be used. The higher the difference between the chosen value and the actual value, the more conservative the behavior of the controller. Once $U_{min}$ is chosen, (\ref{eq:ub}) can be used to choose the suitable upper bounds on velocity and acceleration error estimates.

It is important to note that the estimator gains are selected to balance performance and stability across a range of expected operating conditions.
From (\ref{eq:dc1}), (\ref{eq:res1}), (\ref{eq:11}), and (\ref{eq:ub}), we observe that higher values for $g_1$ and $g_3$ improve estimation accuracy and reduce conservative behavior in continuous-time systems.  Ideally, maximizing $g_1$ and $g_3$ is desirable. However, in practice, these estimator gains should be selected to be as high as possible while considering computational limitations and system response time. The value of $g_2$ should be chosen to ensure a state matrix with negative real eigenvalues, ensuring stability.

\textit{Remark:} To maintain safety, it is necessary to have ${E_v \geq \tilde{v}_1 \; \forall t>0}$. This can be achieved by selecting negative real eigenvalues for the state matrix A. Such eigenvalues guarantee a non-oscillatory convergence to the equilibrium point.

The stability analysis conducted in this section demonstrates that when follower autonomous vehicles are controlled using the proposed safety-critical controller and estimator design, the system maintains stability amidst disturbances and converges to a state where each vehicle follows its predecessor at a fixed, conservative distance. The incorporation of barrier functions in the controller design ensures that this stable state is achieved while operating within the safe region, irrespective of the nonlinear interactions among the estimator state, vehicle dynamics, and inter-vehicular distance. According to Theorem 1, Case 1 addresses the scenario in which the preceding vehicle maintains constant acceleration. Case 2, on the other hand, establishes the system's stability even when the preceding vehicle exhibits non-zero jerk.

\subsection{Adaptive Error Dynamics}\label{sec:aed}
In previous sections, we established that an optimal control law strives to maintain the least conservative distance from the preceding vehicle. Although, according to~(\ref{eq:11}), the conservative bias $h^*$ is proportional to the constant velocity error bound $E_v$ and the estimator gain $g_1$, the system's transient behavior can be further optimized by incorporating adaptive dynamics for the error bound $E_v$. To make the controller adaptive and reduce the conservative bias, we introduce an adaptive state $\epsilon$ (as the velocity error bound) such that $E_v \geq \epsilon \geq \tilde{v}_1$. To enforce these bounds, a new control barrier function, with the safe set $\mathcal{C}_2 \subset \mathbb R^2$ is defined as:
\begin{equation}\label{eq:c12}
    \mathcal{C}_2 = \{ (\epsilon,\tilde{v}_1) \in \mathbb{R}^2 : \eta(\epsilon,\tilde{v}_1) = (E_v - \epsilon)(\epsilon - \tilde{v}_1)\geq 0\}
\end{equation}
Utilizing Definition 1, for any class $\mathcal{K}$ function $\beta$, the system operates within the safety region if, 
$\dot \eta \geq - \beta (\eta)$.
To simplify the notation, let $a = E_v - \epsilon$, $b = \epsilon - \tilde{v}_1$, and $\eta = ab$. Then:
\begin{equation}\label{eq:etadot}
     \dot{\eta} = a (\dot \epsilon - \dot{\tilde{v}}_1 ) - \dot \epsilon b \geq - \beta (\eta)
\end{equation}
We use (\ref{eq:etadot}) to calculate the bounds on $\dot \epsilon$:
\begin{equation}
    \begin{aligned}
     &\dot \epsilon (a-b) \geq a \dot{\tilde{v}}_1 - \beta(ab) \\
     &\dot \epsilon (a-b) \geq a (E_u + g_2 \tilde{d}) - \beta(ab)
    \end{aligned}
\end{equation}
On further simplification, we get the bounds on $\dot \epsilon$ as:
\begin{equation}\label{eq:ed}
     \begin{cases}
         \dot \epsilon \geq \frac{a (E_u + g_2 \tilde{d}) - \beta(ab)}{a-b} \; ; & a > b \\
         \dot \epsilon \leq \frac{a (E_u + g_2 \tilde{d}) - \beta(ab)}{a-b} \; ; & a < b \\
         \dot \epsilon \in \mathbb R \; ; & a = b
     \end{cases}
\end{equation}

Considering the unavailability of $\tilde{v}$ value, any dynamics dependent on $b$ are infeasible. However, the upper and lower bounds on the $\tilde{v}_1$ can be chosen if the initial value is known. Assumption 1 states that $\tilde{v}_1$ is zero at $t=0$. Considering this, let $v_m$ represent the lower bound of $\tilde{v}$ with dynamics as $\dot{v}_m = g_2 \tilde{d} - E_u $ from (\ref{eq:2}). One possible solution dependent on $v_m$, derived in Appendix~\ref{sec:A2}, is presented below, where $\bar \beta$ is the derivative of linear class $\mathcal{K}$ function $\beta$:
\begin{equation}\label{eq:21}
    \begin{cases}
    \dot \epsilon = 0 \; ; & E_u+g_2\tilde{d}<a\bar\beta \\
    \dot \epsilon = a\bar \beta \; ; & E_u+g_2\tilde{d}=a\bar\beta \\
    \dot \epsilon = \frac{a (E_u + g_2 \tilde{d}) - \beta(a (\epsilon - v_m))}{a-(\epsilon - v_m)} \; ; & a > \epsilon - v_m \\
    [\epsilon \;,\; \dot \epsilon] = [E_v \;,\; 0] \; ; & \text{otherwise}
    \end{cases}
\end{equation}

Considering the adaptive law of the velocity error bound in~(\ref{eq:21}) and the dependence of the distance constant $d_c$ on this bound from (\ref{eq:dc1}), we propose that $d_c$ can also be rendered adaptive. The adaptive law for $d_c$ is chosen as follows:
\begin{equation}\label{eq:dc2}
    \dot d_c = \epsilon  + g_1 d_c
\end{equation}
Taking into account the dynamics of $\epsilon$ and $d_c$, the controller design and Lyapunov derivative are subject to change, necessitating re-analysis. Following the calculations from (\ref{eq:8}) and (\ref{eq:hat2}), the control input and conservative function estimate dynamics are obtained as:
\begin{align}
    u &=  \kappa \hat v_1 - \kappa \epsilon - \kappa v + \kappa \alpha (h) \label{eq:22} \\
    \dot{\hat h} &= \epsilon  + g_1 \tilde{d} - \alpha (\hat h - \tilde{d} ) \label{eq:hat3}
\end{align}
The controller formulated in (\ref{eq:22}) offers a continuous, non-singular closed-form solution for cruise control, dependent on the state estimates, sensor measurements, and an adaptive velocity error bound for the follower vehicle, which is calculated by integrating the piece-wise continuous dynamics presented in (\ref{eq:21}).

\textit{Theorem 2:} Given the Lyapunov function in (\ref{eq:V1}), the control law in (\ref{eq:22}), the adaptive law for velocity error bound in (\ref{eq:21}), the conservative distance $d_c$ in (\ref{eq:dc2}), and the conservative function estimate dynamics in (\ref{eq:hat3}), the error estimate $\tilde{e}$ in (\ref{eq:2}) converges to zero and conservative distance $h$ converges to $-\frac{E_v}{g_1}$, for the specific choice of class $\mathcal{K}$ function $\alpha$.

\textit{Proof:} Computing the time derivative of $V_1$ in (\ref{eq:V1}), the following is obtained:
\begin{equation}
    \dot{V}_1 = (\hat h - d_c)(\epsilon  + g_1 \tilde{d} - \alpha (\hat h - \tilde{d} ) - \dot{d}_c) - \tilde{e}^T Q \tilde{e}
\end{equation}
By choosing the class $\mathcal{K}$ function as $\alpha (x) = -g_1x$, we have:
\begin{align}
    \dot{V}_1 = (\hat h - d_c)(\epsilon  + g_1 d_c - \dot{d}_c) - (\hat h - d_c )\alpha (\hat h - d_c ) - \tilde{e}^T Q \tilde{e}
\end{align}
Upon substituting the dynamics of $d_c$ from (\ref{eq:dc2}), the expression for $\dot{V}_1$ is transformed into a negative definite form:
\begin{align}
    \dot{V}_1 = - (\hat h - d_c )\alpha (\hat h - d_c ) - \tilde{e}^T Q \tilde{e} < 0
\end{align}
The conservative distance $d_c$ is incorporated to maintain stability in the presence of estimation error. With $\tilde{v}_1 = 0$ at $t=0$ signifying no initial error, $d_c$ can be initialized as zero. As $\dot{V}_1$ is negative definite, $\tilde{e}$ and $h$ asymptotically converge to zero and $d_c$, respectively. The dynamics in (\ref{eq:21}) indicates that $\epsilon$ converges to $E_v$ as $\tilde{e} \xrightarrow{} 0$. At equilibrium, we have $\dot{d}_c = 0$ and $d_c = -\frac{E_v}{g_1}$, thus $h$ converges to $ -\frac{E_v}{g_1}$.

Theorem 2 asserts that for the proposed adaptive laws of $\epsilon$ and $d_c$, the system states converge to their respective equilibrium points when $u_j = 0$. Similar calculations for $V_2$ from (\ref{eq:V2}) demonstrate that the proposed adaptive laws render the system asymptotically stable for $u_j \neq 0$.

\section{Results}\label{sec:results}
This section showcases the performance and effectiveness of our proposed estimator and control law under various cruise control scenarios. To obtain these results, we employed the following parameters: $g_1\,{=}\,{-9}$, $g_2\,{=}\,{-26}$, $g_3\,{=}\,{-24}$, $U_{min}=$ ${-0.923}$ m/s$^3$, $E_v\,{=}\,0.346$ m/s, $E_u\,{=}\,1$ m/s$^2$, and $\beta (x)\,{=}\,x$. The state matrix $A$ exhibits Hurwitz properties with these values, featuring eigenvalues of $[{-2},{-3},{-4}]$. To demonstrate the robustness and efficacy of the proposed system, the following scenarios are examined: a) acceleration and deceleration of the leader vehicle; b) propagation of errors in the network (string stability); c) performance enhancement through adaptive control law design proposed in Section~\ref{sec:aed}; and d) real-world performance evaluation for an acceleration-deceleration scenario in physics-realistic AirSim environment~\cite{airsim23}.

The AirSim environment, with its sophisticated physics engine, is utuilized for simulating real-world scenarios in part (d), due to its capabilities to mimic the behavior and conditions a vehicle might encounter in actual operation, such as dynamic interactions with other vehicles. For the evaluation scenarios in parts (a), (b), and (c), MATLAB serves as a suitable simulation tool, adept at generating precise and comprehensive results. The continuous-time behavior of the system was achieved using the 4th-order Runge-Kutta method. Moreover, the difference in states between vehicles was used to simulate the distance sensor data for each vehicle. This computational approach, combined with the chosen scenarios in both MATLAB and AirSim, facilitates a comprehensive examination of the proposed system's behavior under diverse conditions, providing valuable insights into its performance and effectiveness in real-world applications.

\subsection{Acceleration -- Deceleration Scenarios}
\begin{figure}[t]
    \centering
    \includegraphics[width=\columnwidth]{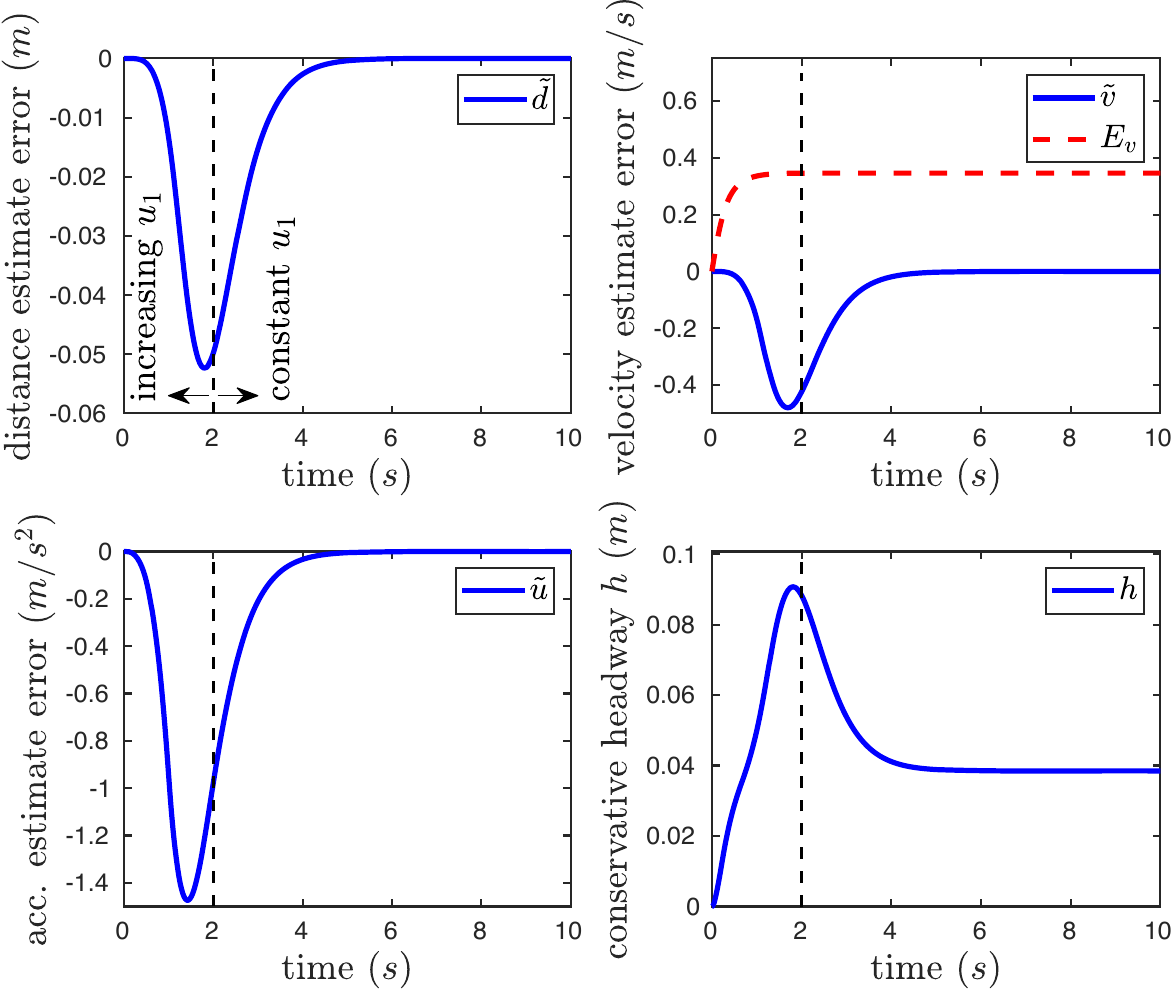}
    \caption{Performance of the estimator and conservative behavior of the control law when leader vehicle accelerates.}
    \label{fig:fig3}
\end{figure}

In the acceleration scenario, the lead vehicle is programmed to undergo rapidly increasing acceleration (i.e., positive jerk) for two seconds, then maintaining a constant acceleration for the remainder of the experiment.
Starting from rest conditions with $\|\tilde{d}\|=\|\tilde{v}\|=\|\tilde{u}\|=0$ and an ideal headway conservative distance $h = 0$, the estimator error is expected to converge to the origin, as the headway distance approaches the conservative value $d_c\,{=}\,\nicefrac{-E_v}{g_1}\,{=}\,0.0384$ m for this particular trial. As illustrated in Fig.~\ref{fig:fig3}, the estimator error initially increases due to the lead vehicle's rapid acceleration but swiftly converges to the origin once the vehicle sustains constant acceleration. It is important to note that maintaining a headway distance of $h=0$ m is infeasible as the ACC system operates without communication. The Lyapunov function's design compels the system to preserve a conservative distance, which also contributes to the deviation of the estimator error from its initial value.

\begin{figure}[t!]
    \centering
    \includegraphics[width=\columnwidth]{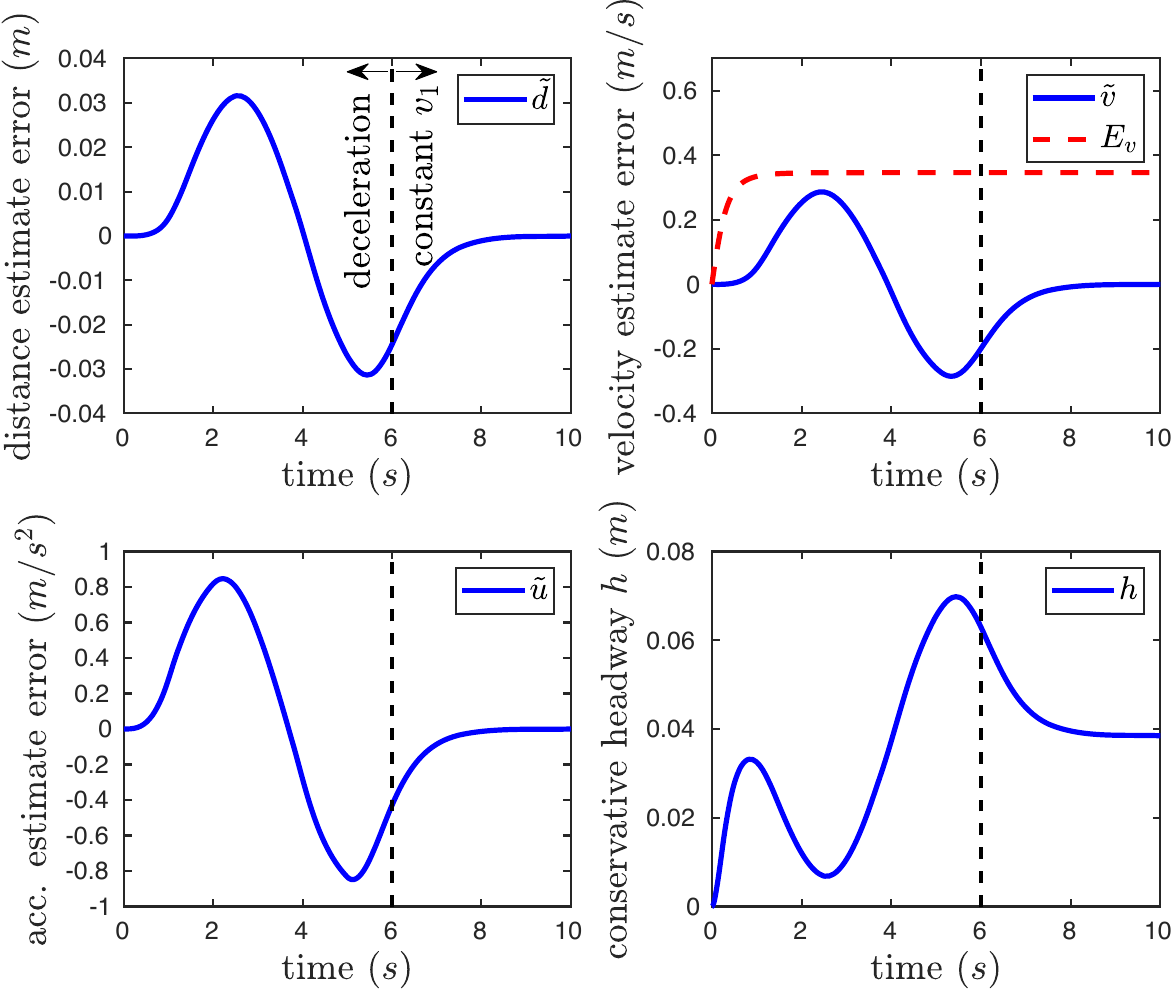}
    \caption{Performance of the estimator and conservative behavior of the control law when leader vehicle decelerates.}
    \label{fig:fig4}
\end{figure}

In the deceleration scenario, which closely resembles the acceleration scenario, the lead vehicle is programmed to decelerate for six seconds while adhering to the constraints defined by (\ref{eq:ub}), subsequently, the vehicle maintains a constant velocity.
The system starts with communication and ideal headway distance conditions.
As demonstrated in Fig.~\ref{fig:fig4}, when the lead vehicle brakes, the estimator quickly converges to the origin while ensuring safety ($h>0$).

\begin{figure}[t]
    \centering
    \includegraphics[width=\columnwidth]{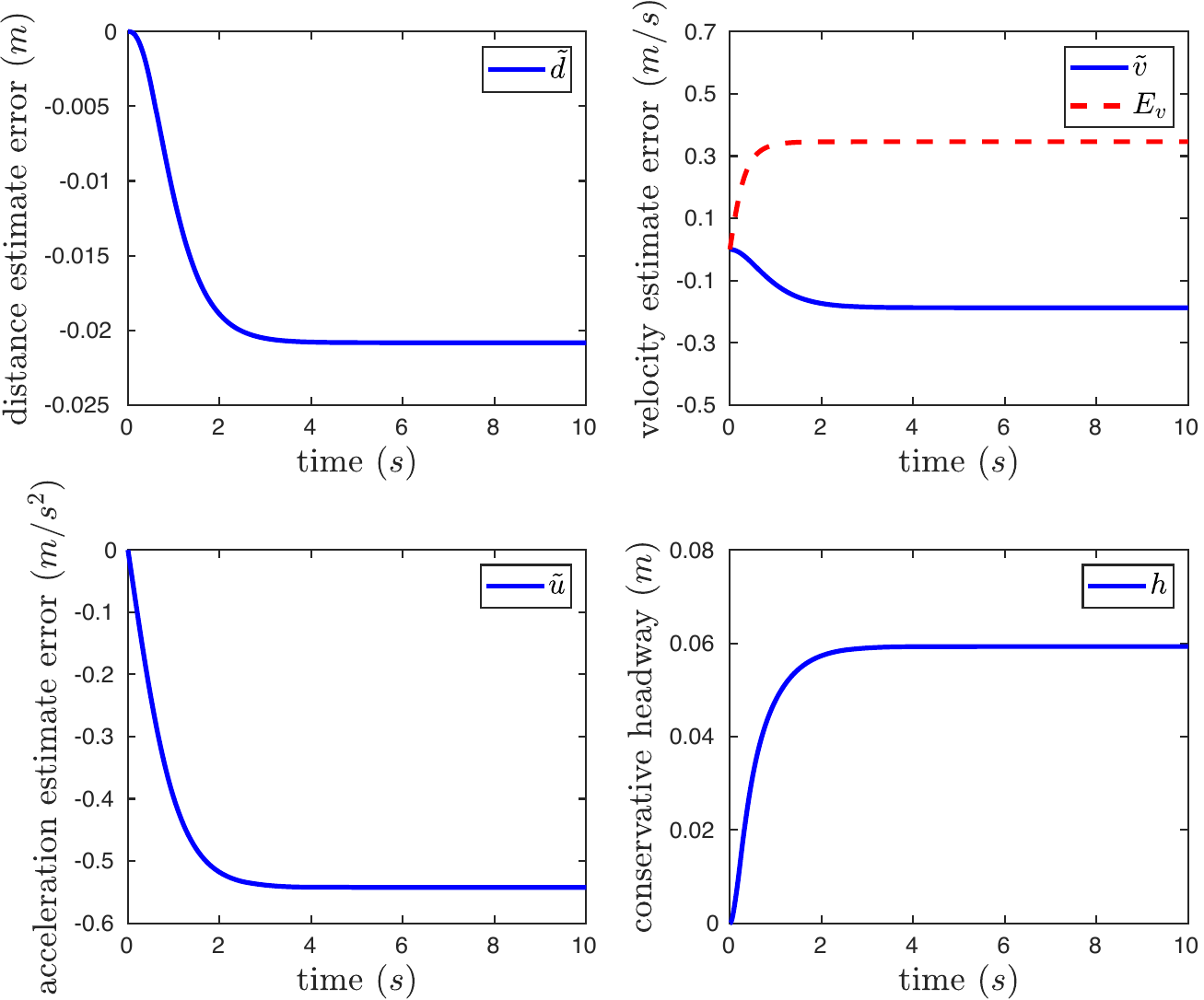}
    \caption{Performance of the estimator and conservative behavior of the control law for constant leader vehicle jerk $u_j$.}
    \label{fig:fig5}
\end{figure}

Another important scenario involves the lead vehicle undergoing constant acceleration or deceleration with a fixed jerk ($u_j$). Fig.~\ref{fig:fig5} displays the system's performance when subjected to $u_j\,{=}\,0.5$ m/s$^3$, starting from rest and with an ideal headway distance. As  derived in (\ref{eq:res1}), the system converges to $\tilde{d}^* = 0.0208$ m, $\tilde{v}^* = 0.187$ m/s, $\tilde{u}^* = 0.5416$ m/s$^2$, and $h^* = 0.593$ m.

It is crucial to note that, for all the scenarios discussed in this section, the results show that the velocity and acceleration estimation errors satisfy $E_v \geq \tilde{v}_1$ and $E_u \geq \tilde{u}_1$ even under extreme conditions. Furthermore, while all the presented scenarios commence with an ideal headway distance, this is not a requirement. Any headway greater than the ideal distance has been proven to converge.

\subsection{String Stability Performance}
\begin{figure}[t]
    \centering
    \begin{tikzpicture}
    \node (img) {\includegraphics[width=\columnwidth]{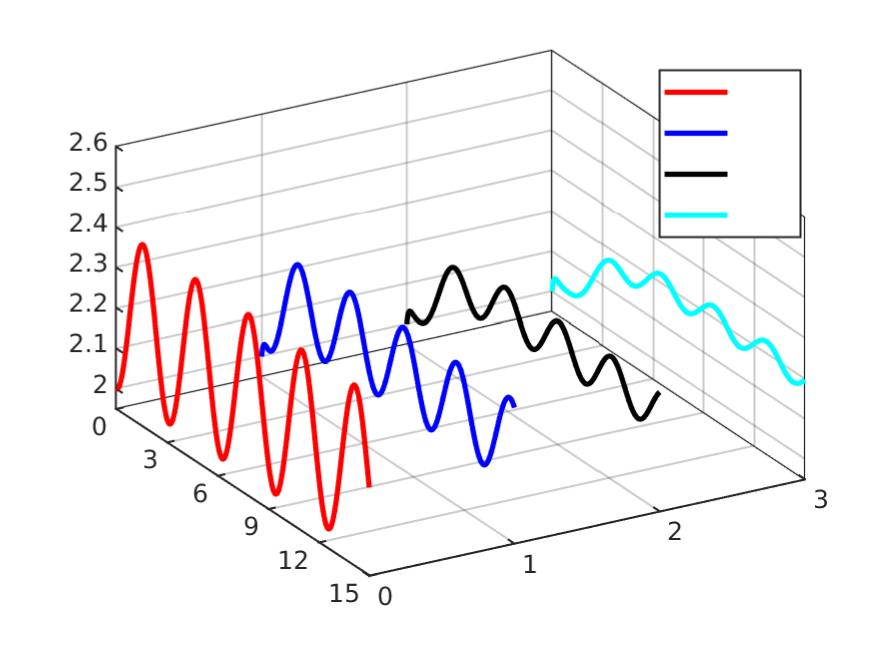}};
    \node [align=center] at (-2.75,-2.3){\footnotesize time (s)};
    \node [align=center] at (3.15,2.31){\footnotesize LV};
    \node [align=center] at (3.15,1.91){\footnotesize AV1};
    \node [align=center] at (3.15,1.51){\footnotesize AV2};
    \node [align=center] at (3.15,1.11){\footnotesize AV3};
    \node [align=center,rotate=90] at (-4.15,0.45){\footnotesize vehicle velocity $(m/s)$};
    \end{tikzpicture}
    \vspace{-25pt}
    \caption{String stability in a four-vehicle platoon under ACC framework.}
    \label{fig:fig6}
\end{figure}

\begin{figure}[t]
    \centering
    \begin{tikzpicture}
    \node (img) {\includegraphics[width=\columnwidth]{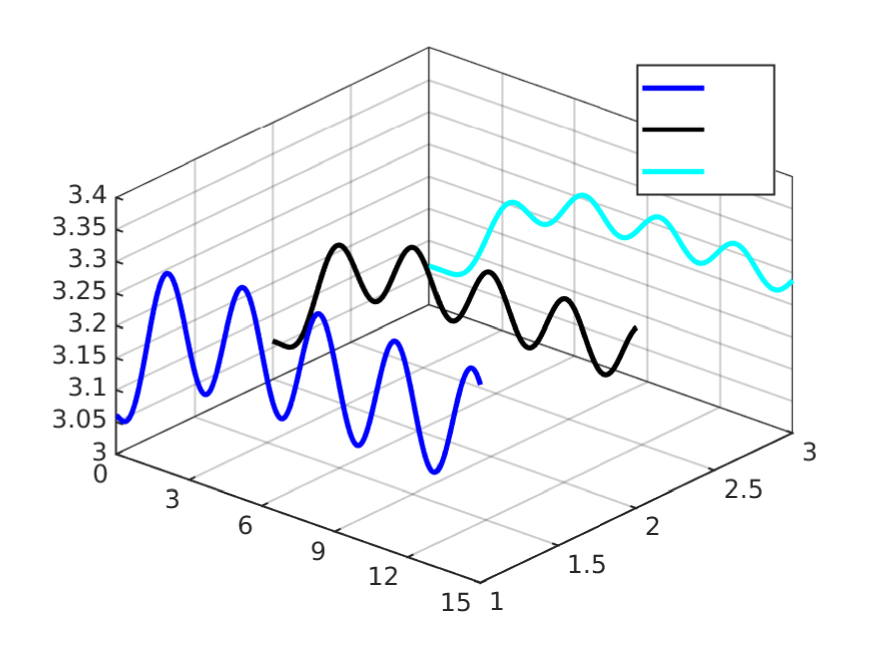}};
    \node [align=center] at (-2.55,-2.5){\small time (s)};
    \node [align=center] at (3.05,2.45){\footnotesize AV1};
    \node [align=center] at (3.05,2.02){\footnotesize AV2};
    \node [align=center] at (3.05,1.60){\footnotesize AV3};
    \node [align=center,rotate=90] at (-4.25,0){\small inter-vehicle distance $(m)$};
    \end{tikzpicture}
    \vspace{-25pt}
    \caption{Inter-vehicle distance in a four-vehicle platoon under ACC framework.}
    \label{fig:fig7}
\end{figure}

String stability is a critical performance measure for ACC systems. It ensures that disturbances like sudden braking or acceleration do not propagate through the platoon, causing potential accidents or traffic congestion.
In a string-stable ACC system, the amplitude of the sinusoidal velocity should decrease as it progresses from the lead vehicle to the end of the platoon. Fig.~\ref{fig:fig6} and Fig.~\ref{fig:fig7} demonstrate the performance of the proposed control law design in a four-vehicle platoon consisting of one lead vehicle and three autonomous vehicles. 
In this experiment, the leader vehicle's velocity was disturbed by an amplitude of $0.2$ m/sec. As this disturbance propagated upstream through the platoon, the amplitude decreased to $0.1068$ m/sec, $0.0609$ m/sec, and $0.0369$ m/sec for AV1, AV2, and AV3, respectively. This results in an average string stability gain ($\mathcal{S}$) of $0.57$. The reduced amplitude of the velocity profiles (in comparison to the lead vehicle) towards the end of the platoon, along with $\mathcal{S}<1$, confirms the string stability of the control law. 
The designed controller inherently ensures string stability, and the magnitude of the string stability gain can be adjusted by the eigenvalues of the estimator's state matrix $A$.

\subsection{Connected Cruise Control}
Connected cruise control is an equivalent of ACC, with the added capability for vehicles to communicate their states. ACC designs are communication-independent and usually exhibit no performance improvements when employed in CCC applications, making them equivalent to scenarios without communication. However, estimator-based control laws can utilize communication to initialize estimator states, and the control design based on these estimated states should ideally perform equal to or better than the no-communication scenario.

The performance of the proposed control law depends on the upper bound of the estimated error and can demonstrate improvements when this upper bound is reduced. In the absence of communication noise and delay, communication with the lead vehicle is equivalent to a zero upper bound on the estimated error. Integrating this with the adaptive error dynamics proposed in Section~\ref{sec:aed} is expected to enhance the performance of the CCC system.

\begin{figure}[t!]
    \centering
    \setlength{\tabcolsep}{-2.5pt}
    \renewcommand{\arraystretch}{0.8}
    \begin{tabular}{cc}
    \begin{tikzpicture}
    \node (img) {\includegraphics[width=0.49\columnwidth]{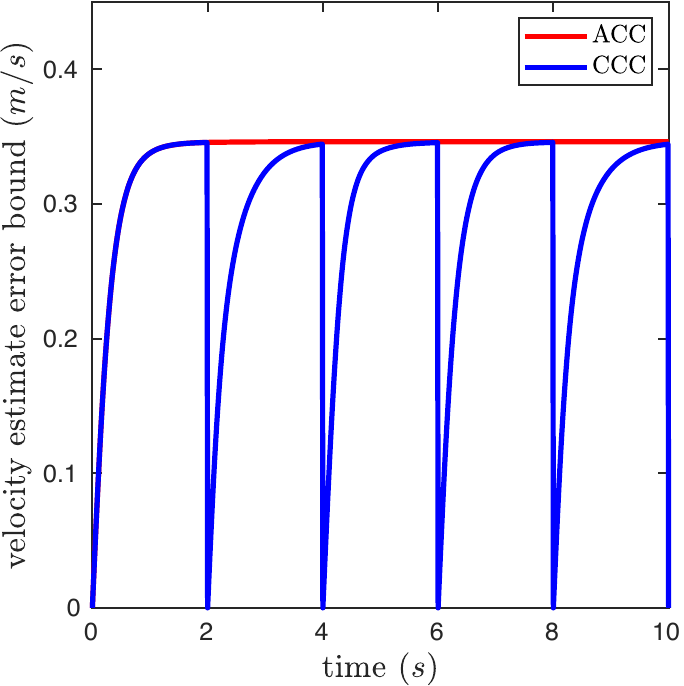}};
    \node [align=center,fill=white] at (0.25,-2.2){\footnotesize time (s)};
    \node [align=center,rotate=90,fill=white] at (-2.15,0.05){\footnotesize velocity estimate error bound $(m/s)$};
    \end{tikzpicture} & 
    \begin{tikzpicture}
    \node (img) {\includegraphics[width=0.49\columnwidth]{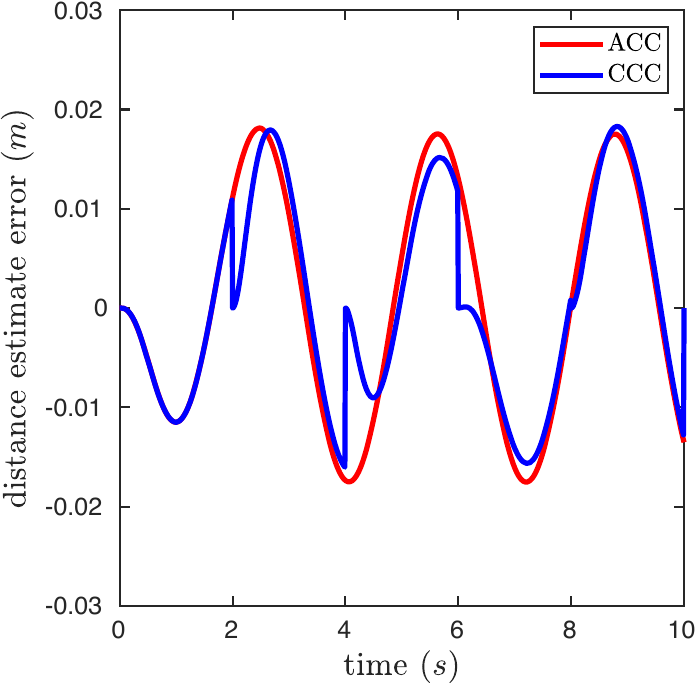}};
    \node [align=center,fill=white] at (0.3,-2.15){\footnotesize time (s)};
    \node [align=center,rotate=90,fill=white] at (-2.15,0.175){\footnotesize distance estimate error $(m)$};
    \end{tikzpicture}
    \tabularnewline
    \small{(a)} & \small{(b)}
    \tabularnewline
    \begin{tikzpicture}
    \node (img) {\includegraphics[width=0.49\columnwidth]{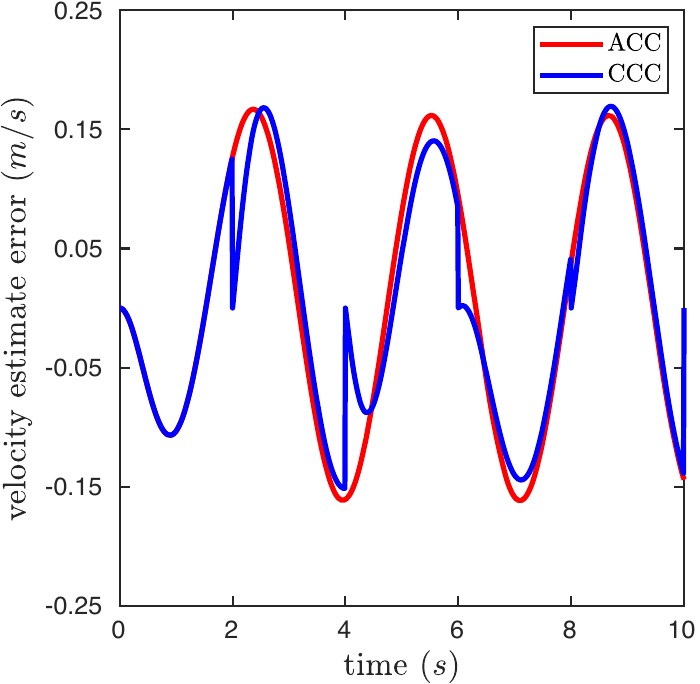}};
    \node [align=center,fill=white] at (0.25,-2.15){\footnotesize time (s)};
    \node [align=center,rotate=90,fill=white] at (-2.15,0.1){\footnotesize velocity estimate error $(m/s)$};
    \end{tikzpicture} & 
    \begin{tikzpicture}
    \node (img) {\includegraphics[width=0.49\columnwidth]{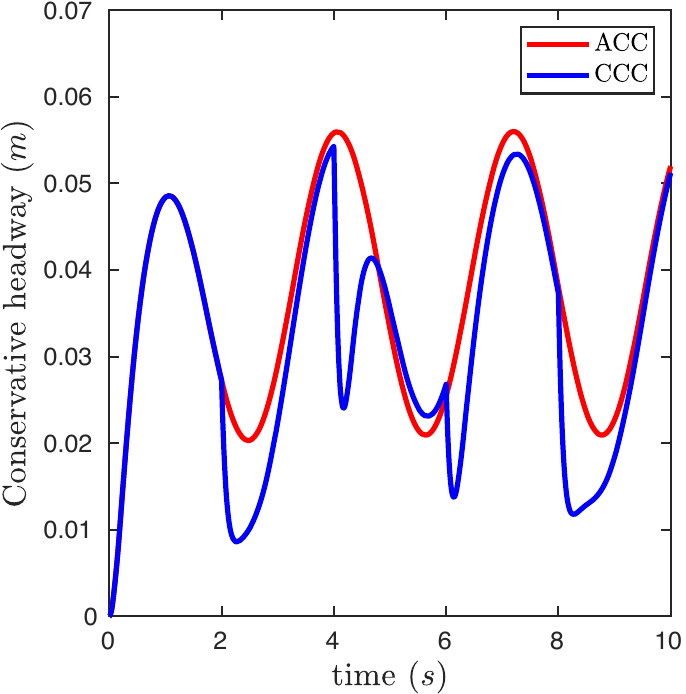}};
    \node [align=center,fill=white] at (0.3,-2.15){\footnotesize time (s)};
    \node [align=center,rotate=90,fill=white] at (-2.15,0.175){\footnotesize conservative headway $(m)$};
    \end{tikzpicture}
    \tabularnewline
    \small{(c)} & \small{(d)}
    \end{tabular}
    \caption{Comparison of system performance metrics between ACC and CCC systems: (a) velocity estimation error bound, (b) distance estimation error, (c) velocity estimation error, and (d) conservative headway distance. The CCC system communicates every 2 seconds.}
    \label{fig:fig8}
\end{figure}

The system’s performance was compared under two conditions: with V2V communication (CCC) and without it (ACC) when the leader vehicle experienced sinusoidal acceleration $u_j$. Fig.~\ref{fig:fig8} presents a comparison of several parameters, including velocity estimation error bounds, distance estimation error, velocity estimation error, and conservative headway for both scenarios.
As shown in Fig.~\ref{fig:fig8}(a), in the CCC scenario, the estimation error drops to zero following each communication instance and then gradually increases up to the maximum error bound. In the ACC scenario, the error gradually increases to the maximum error bound and remains constant despite the absence of communication.
The performance of the estimator depends on the communicated states and their incorporation into the design. Specifically, the CCC scenario under consideration assumes communication of vehicle velocities and the re-initialization of the estimator state with these communicated values. Fig.~\ref{fig:fig8}(b) and (c) demonstrate that, compared to the ACC case, the overall estimation error is lower with CCC.

The effectiveness of a cruise control system is indicated by the conservative headway maintained between vehicles. The proposed controller design, which utilizes velocity estimation error bounds and estimator states, aims to maintain the least safe headway from preceding vehicles. Consequently, more accurate estimates and smaller error bounds are likely to result in less conservative behavior. Fig.~\ref{fig:fig8}(d) depicts the time headway's evolution, highlighting the reduced conservative headway with adaptive error dynamics in the CCC scenario compared to the constant error bound in the ACC scenario.

Table~\ref{tab:table1} summarizes the averages of estimation error bound, absolute distance estimation error, absolute velocity estimation error, and conservative headway for ACC and CCC scenarios.  The average conservative headway maintained in the CCC system is 0.0317m, slightly surpassing the 0.0364m maintained in the ACC scenario. The conservativeness of the CCC scenario is marginally less compared to the ACC scenario, as the average estimation error bound is closer to zero for CCC. Consequently, the averages of absolute distance and velocity estimation errors are also smaller in the CCC scenario.
This performance difference between the ACC and CCC scenarios becomes more distinct with higher speeds and larger platoons, underlining the overall effectiveness of both systems in ensuring safe and efficient vehicle control, with CCC demonstrating a slight edge in maintaining a more precise and marginally smaller conservative headway.

\begin{table}[t]
    \caption{Magnitude of system performance metric \\ for ACC and CCC systems}
    \label{tab:table1}
    \centering
    \begin{tabular}{|c|c|c|} 
    \hline
    Performance Metric & ACC & CCC \\ 
    \hline
    Avg. estimation error bound & 0.3363 & 0.2932 \\ 
    Avg. of absolute distance estimation error & 0.0102 & 0.0086 \\
    Avg. of absolute velocity estimation error & 0.0957 & 0.0841 \\ 
    Avg. conservative headway & 0.0364 & 0.0317 \\ 
    \hline
    \end{tabular}
\end{table}

\subsection{Real-World Performance Evaluation}

\begin{figure}[b!]
    \centering
    \begin{tikzpicture}
    \node (img) {\includegraphics[width=0.35\textwidth]{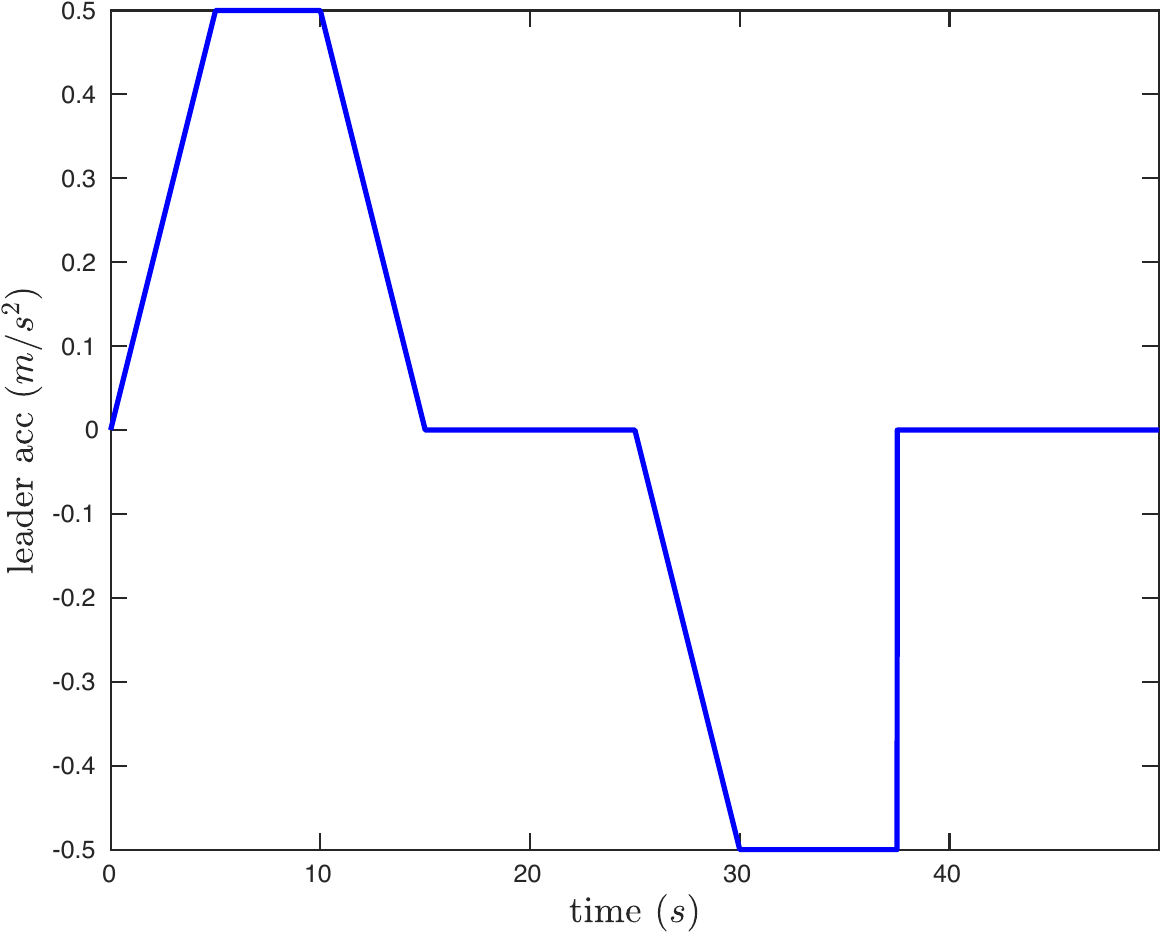}};
    \node [align=center,fill=white] at (0.25,-2.55){\footnotesize time (s)};
    \node [align=center,rotate=90,fill=white] at (-3.2,0.3){\footnotesize leader acceleration $(m/s^2)$};
    \end{tikzpicture}
    \vspace{-5pt}
    \caption{Leader acceleration profile for the real-world performance evaluation.}
    \label{fig:fig9}
\end{figure}

\begin{figure*}[t]
    \renewcommand{\arraystretch}{0.5}
    \setlength\tabcolsep{0pt}
    \centering
    \begin{tabular}{ccc}
    \begin{tikzpicture}
    \node (img) {\includegraphics[width=0.325\textwidth]{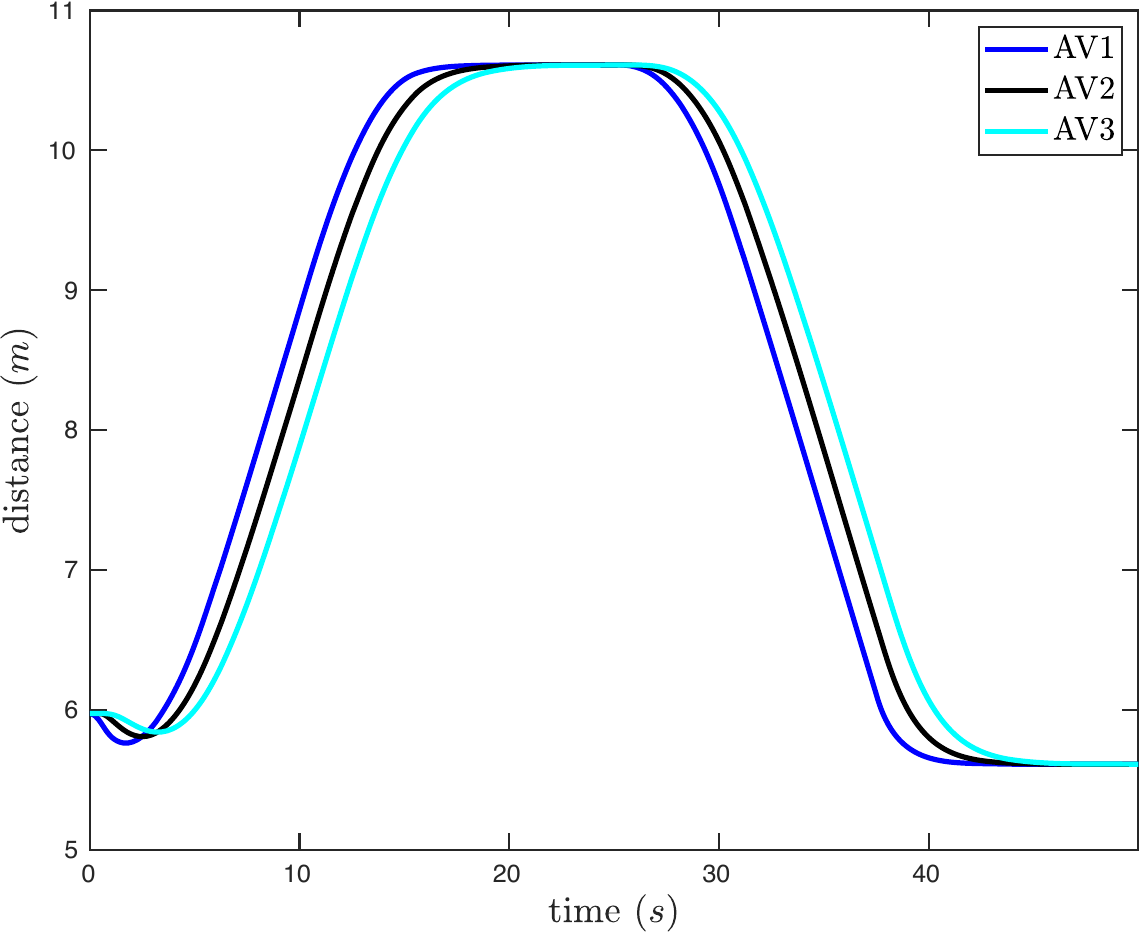}};
    \node [align=center,fill=white] at (0.25,-2.4){\footnotesize time (s)};
    \node [align=center,rotate=90,fill=white] at (-2.95,0.2){\footnotesize distance $(m)$};
    \end{tikzpicture} & 
    \begin{tikzpicture}
    \node (img) {\includegraphics[width=0.33\textwidth]{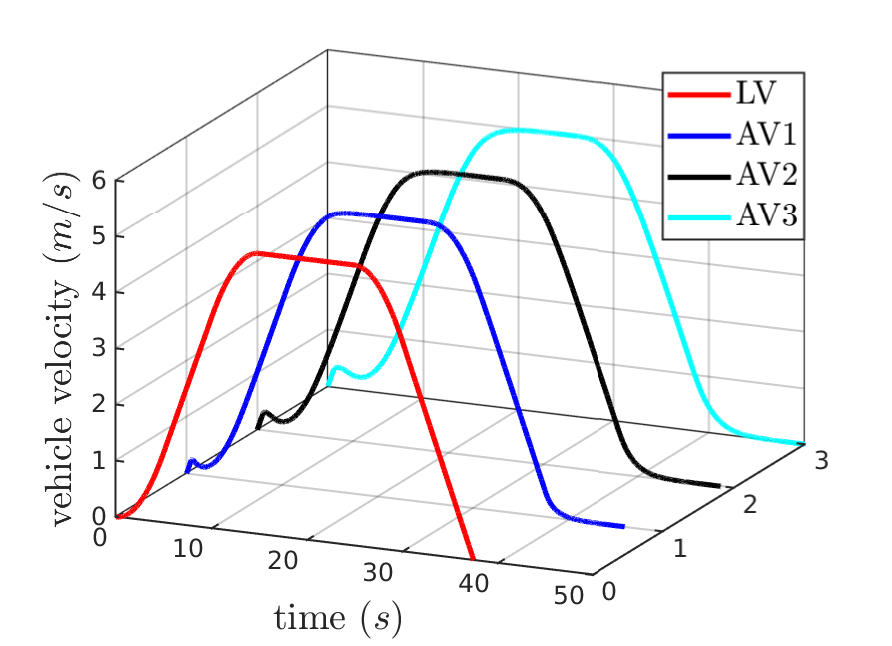}};
    \node [align=center,fill=white] at (-0.7,-2.0){\footnotesize time (s)};
    \node [align=center,rotate=90,fill=white] at (-2.65,-0.1){\footnotesize vehicle velocity $(m/s)$};
    \end{tikzpicture} & 
    \begin{tikzpicture}
    \node (img) {\includegraphics[width=0.33\textwidth]{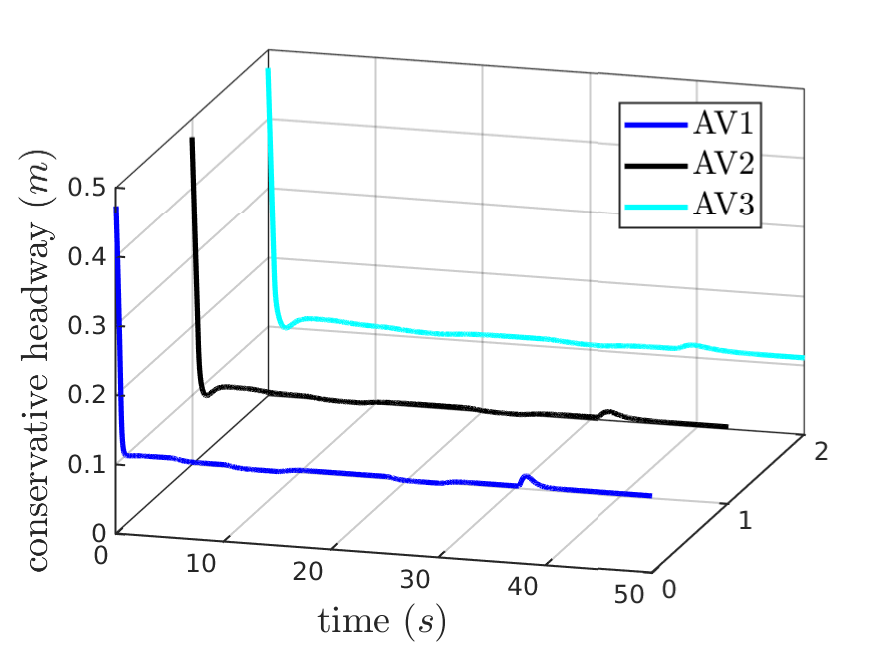}};
    \node [align=center,fill=white] at (-0.5,-2.1){\footnotesize time (s)};
    \node [align=center,rotate=90,fill=white] at (-2.8,-0.1){\footnotesize conservative headway $(m)$};
    \end{tikzpicture}
    \tabularnewline
    \small{(a)} & \small{(b)} & \small{(c)}
    \tabularnewline
    \begin{tikzpicture}
    \node (img) {\includegraphics[width=0.325\textwidth]{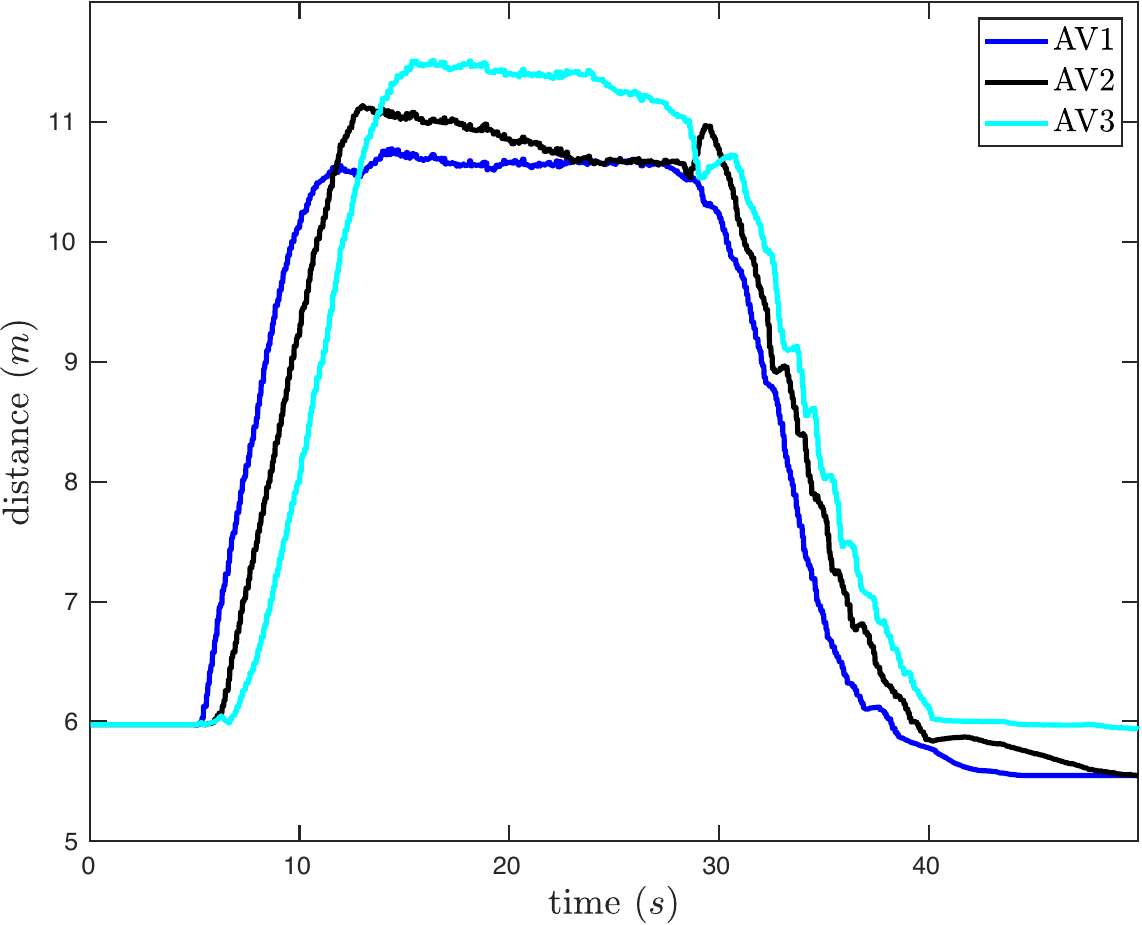}};
    \node [align=center,fill=white] at (0.25,-2.4){\footnotesize time (s)};
    \node [align=center,rotate=90,fill=white] at (-2.95,0.2){\footnotesize distance $(m)$};
    \end{tikzpicture} & 
    \begin{tikzpicture}
    \node (img) {\includegraphics[width=0.33\textwidth]{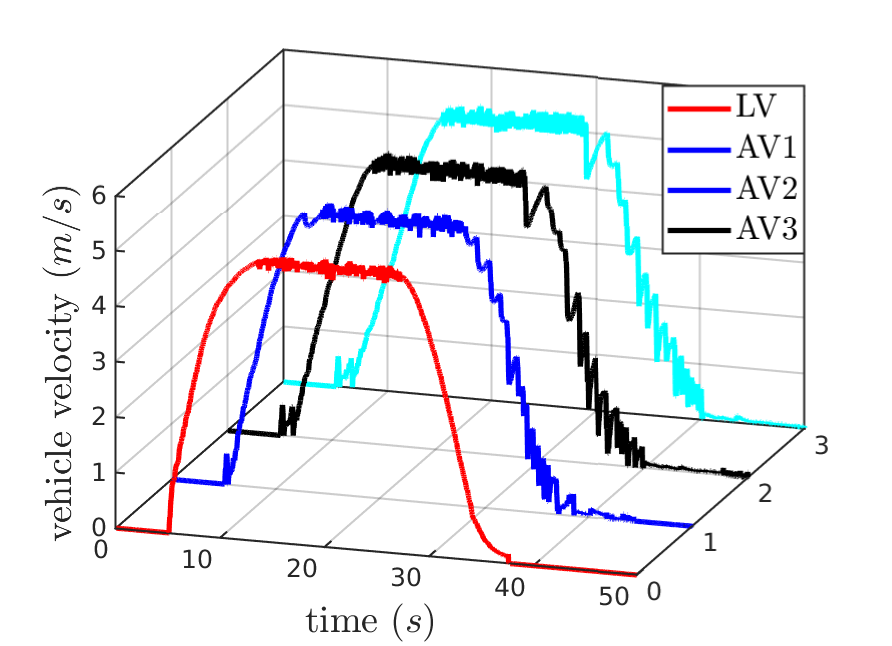}};
    \node [align=center,fill=white] at (-0.6,-2.0){\footnotesize time (s)};
    \node [align=center,rotate=90,fill=white] at (-2.65,-0.1){\footnotesize vehicle velocity $(m/s)$};
    \end{tikzpicture} & 
    \begin{tikzpicture}
    \node (img) {\includegraphics[width=0.33\textwidth]{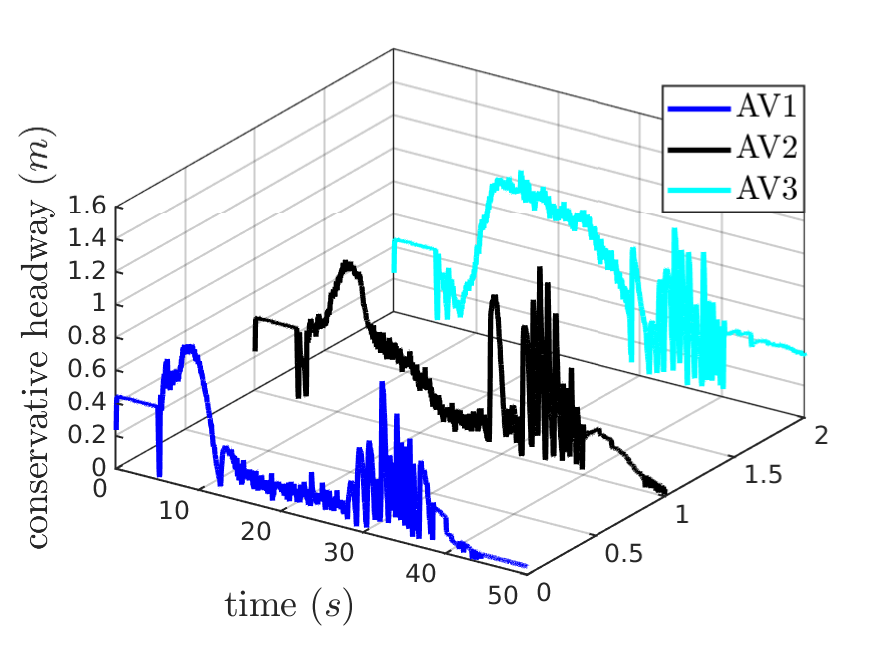}};
    \node [align=center,fill=white] at (-0.95,-1.9){\footnotesize time (s)};
    \node [align=center,rotate=90,fill=white] at (-2.85,-0.1){\footnotesize conservative headway $(m)$};
    \end{tikzpicture}
    \tabularnewline
    \small{(d)} & \small{(e)} & \small{(f)}
    \end{tabular}
    \caption{Evolution of distance, vehicle velocity, and conservative headway over time when leader vehicle follows the acceleration profile shown in Fig.~\ref{fig:fig9} in MATLAB (a, b, c) and AirSim (d, e, f) environments.}
    \label{fig:fig10}
\end{figure*}

In order to assess the feasibility and verify the performance and efficacy of the proposed estimation-based controller in real-world scenarios, we conduct evaluations within AirSim, a physically accurate simulation environment. This section presents the dynamic evolution of various system states for a specific scenario in both MATLAB and AirSim simulators and highlights the similarities and differences between the results of the two platforms.
As an open-source platform, AirSim addresses the need to bridge the gap between simulation and reality, offering a powerful tool for autonomous vehicle development. AirSim provides high-fidelity physical and visual simulation, enabling the development of algorithms and controllers within the simulator and their seamless deployment on vehicles without requiring modifications. It includes essential components such as environment and vehicle models, a physics engine, sensor models, rendering interfaces, public APIs, and an interface layer for vehicle firmware. Leveraging the power of the Unreal Engine~\cite{ue4}, it emulates sensors like vision cameras and LiDARs, providing a realistic perception environment. The vehicle model in AirSim encompasses essential dynamics parameters such as mass, inertia, drag coefficients, friction coefficients, and restitution, allowing the physics engine to compute rigid body dynamics accurately. Additionally, the environment model incorporates crucial aspects such as gravity, air density, air pressure, and the magnetic field. The physics engine incorporated in AirSim enables real-time hardware-in-the-loop (HITL) simulations, operating at a high frequency to accurately emulate vehicle dynamics. Overall, AirSim's combination of high-fidelity simulation, seamless integration with actual vehicles, and rich feature set make it an excellent choice for testing autonomous vehicle control algorithms~\cite{airsim23}.

We analyze a platoon of four vehicles operating in the absence of inter-vehicle communication. The platoon is structured such that the lead vehicle drives with the acceleration profile illustrated in Fig.~\ref{fig:fig9}, and the three other autonomous vehicles operate under the proposed controller. The platoon starts at rest, with all vehicles separated by a distance of 6 m. 
AirSim neighborhood environment is used with all vehicles using a custom low-level controller to match the desired throttle for the corresponding input accelerations. 
The acceleration profile shown in Fig.~\ref{fig:fig9} mimics the acceleration curve typically encountered when navigating urban city traffic, characterized by slow response times and aggressive braking commonly exhibited by human drivers. These conditions frequently lead to traffic congestion. Consequently, this scenario offers a practical and relevant context for the application of ACC systems.
Additionally, the vehicles are equipped with LiDAR sensors to measure the distance from the vehicle or obstacle ahead.
These LiDAR sensors are intentionally modeled in AirSim to include realistic noise in the input data. Introducing noise into the sensor data is essential, as it imitates the imperfections and uncertainties typically encountered in real-world applications. Operating under such noisy sensors enables a thorough evaluation of the design under conditions that closely resemble practical, real-world scenarios.

\begin{figure*}[ht!]
    \centering
    \includegraphics[width=\textwidth]{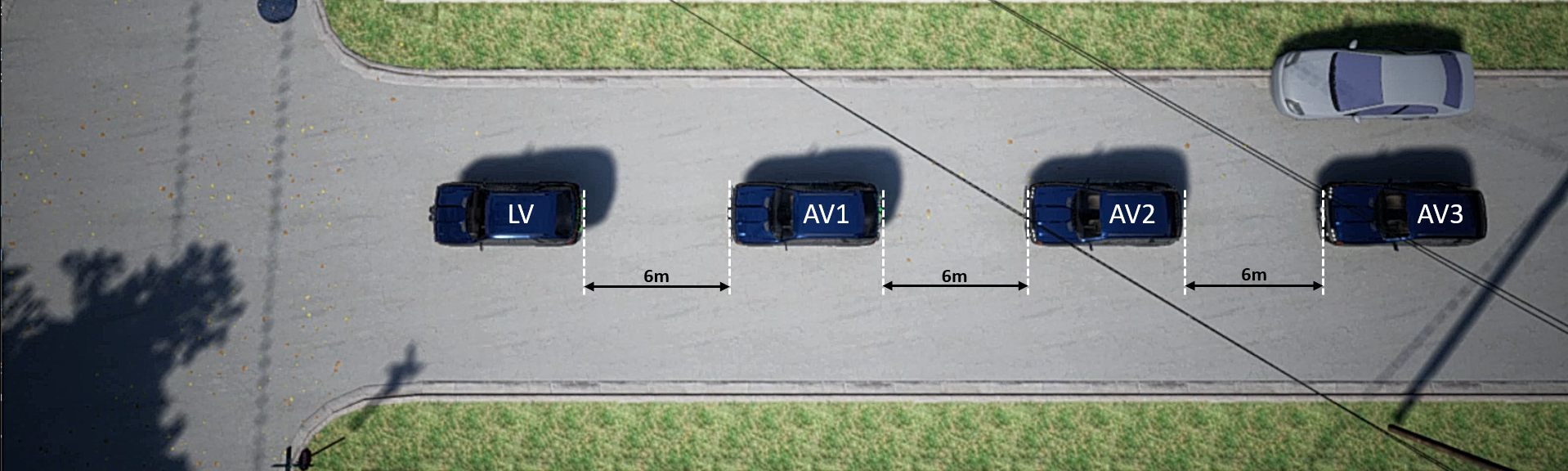} \\
    {(a) $t = 0$ sec.}
    \includegraphics[width=\textwidth]{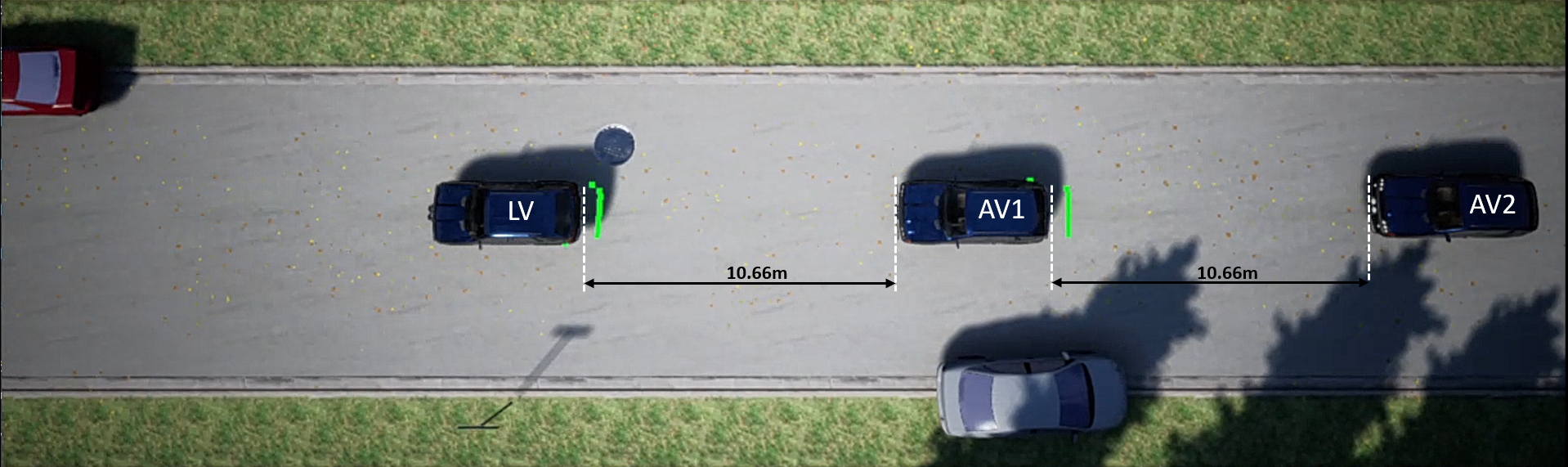} \\
    {(b) $t = 25$ sec.}
    \includegraphics[width=\textwidth]{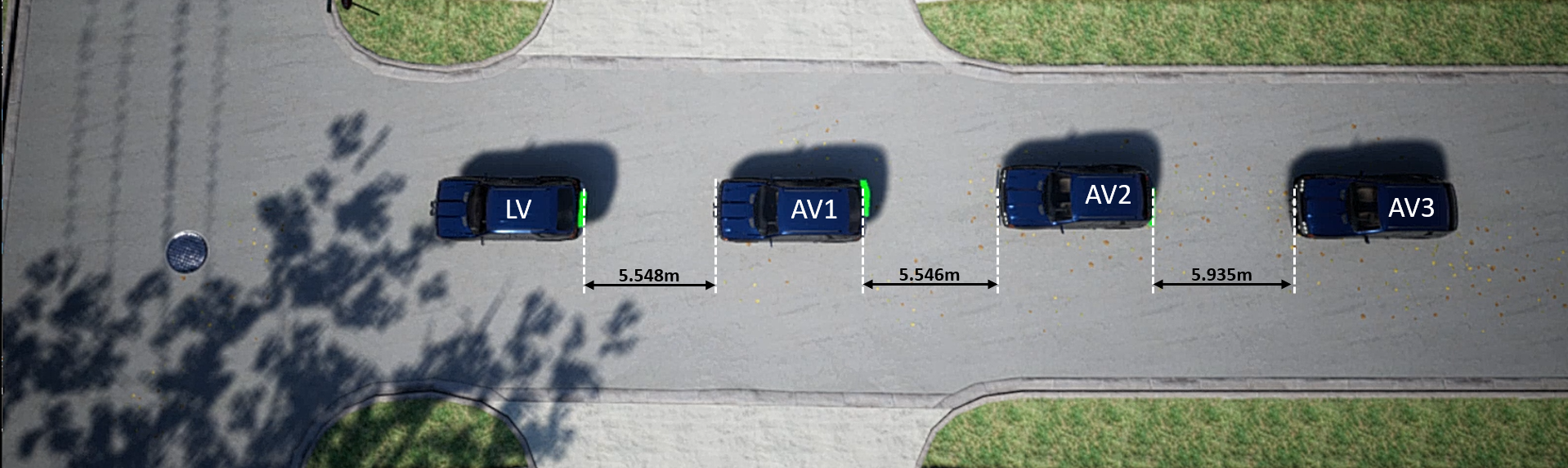} \\
    {(c) $t = 50$ sec.}
    \caption{Snapshot of the AirSim environment showing: (a) the start of the experiment, (b) the middle of the experiment, and (c) the end of the experiment.}
    \label{fig:fig11}
\end{figure*}

The snapshot of the AirSim environment at different timestamps throughout the experiment is presented in Fig.~\ref{fig:fig11}.
Fig.~\ref{fig:fig11}(a) shows the starting state of the experiment with all the vehicles at rest separated by a distance of $6$ m. The state of the experiment at $25$ sec, when all the vehicles are at a speed of about $5$ m/sec and are separated by a distance more than $10.6$ m, is shown in Fig.~\ref{fig:fig11}(b). The end of the experiment, when all the vehicles come to rest, is displayed in Fig.~\ref{fig:fig11}(c). 
In the simulation snapshots in Fig.~\ref{fig:fig11}(b,c), green marks indicate LiDAR-detected objects or obstacles, which are the points reflecting the light from LiDAR. This data is needed for the system's environment perception, enabling real-time performance to maintain a safe following distance.
The snapshots also feature distance markings and vehicle labels.

The parameters employed for the controller and estimator are $g_1\,{=}\,{-9}$, $g_2\,{=}\,{-26}$, $g_3\,{=}\,{-24}$, $E_v\,{=}\,1.0$ m/s, $E_u\,{=}\,1.0$ m/s$^2$, and $\beta (x)\,{=}\,x$. The state matrix $A$ exhibits Hurwitz properties with these values, featuring eigenvalues of $[{-2},{-3},{-4}]$.

Upon performing the calculations, all the autonomous vehicles are expected to stop at a distance of $5.612$ m from the preceding vehicle while maintaining a conservative headway $h>0$ throughout the experiment. The evolution of key states, such as distance, vehicle velocity, and conservative headway is presented in Fig.~\ref{fig:fig10} (a,b,c) for MATLAB simulation with the system dynamics described in (\ref{dyn:1}). It is essential to note that this dynamics model does not account for disturbances arising from factors such as friction, air drag, and noise sensor input. Therefore the obtained results align with the calculations and are considered an ideal response. The AirSim environment model and physics engine incorporate external and internal factors, resulting in deviations from the ideal (MATLAB) response. Fig.~\ref{fig:fig10}(d) shows the distance readings recorded by the LiDAR of autonomous vehicles. Due to real-world factors like friction in AirSim, vehicles cannot maintain very low speeds, preventing them from achieving the desired safe distance from the preceding vehicle. This difference is evident from the plots in Fig.~\ref{fig:fig10}(a), where, at both the start and end of the experiment, the vehicle models in MATLAB manage to maintain very low velocities to achieve distances close to safe distance limit $d_r$.
The autonomous vehicles AV1, AV2, and AV3 in MATLAB converge to $5.6102$ m, $5.6103$ m and $5.6105$ m, respectively, achieving the desired distance with a 2mm error. This error is negligible and is caused by integration errors in the simulations.
However, attaining such small velocities is unfeasible for vehicles in the AirSim environment, leading AV1, AV2, and AV3 to stop at $5.548$ m, $5.5465$ m, and $5.9357$ m, respectively, as shown in Fig.~\ref{fig:fig10}(d).
MATLAB vehicle model employs a continuous time controller, yielding smooth velocity curves as depicted in Fig.~\ref{fig:fig10}(b). In contrast, the AirSim vehicle models implement a discrete-time controller derived from the proposed continuous time controller. The inherent discretization of the controller, combined with the influence of various environmental factors, introduces disturbances into the system, evident in Fig.~\ref{fig:fig10}(e). However, emphasizing safety as the most critical requirement in cruise control applications, Fig.~\ref{fig:fig10}(f) validates the controller's ability to maintain safety standards despite the presence of noise and disturbance in the real-world system. The minimum headway recorded for the three autonomous vehicles is $0.0125$ m, $0.0104$ m, and $0.0155$ m, respectively, indicating that the system successfully delivers a conservative response. The conservative headway for the three autonomous vehicles converges to $0.0480$ m, $0.0465$ m, and $0.4357$ m, respectively, at the end of the experiment, compared to $0.1$ m for MATLAB. Accordingly, it is possible to further improve the conservative response of the controller towards different factors by incorporating them in the dynamics and following the calculations as presented in Section~\ref{sec:controller}.

Videos of the experiments presented in this section can be found in the accompanying video for the article, available at \\ \texttt{\url{https://vimeo.com/849612069/60563a96f7}}

The results obtained through the AirSim simulation experiment substantiate the applicability and practical viability of the proposed controller in real-world scenarios. The comprehensive evaluation demonstrates the controller's safety-critical performance and seamless compatibility with low-level controllers, underscoring the controller's potential for deployment in adaptive cruise control applications. 

\section{Conclusions and Future Directions}\label{sec:conc}

In this manuscript, we have designed an estimator and the associated safety-critical controller for Adaptive Cruise Control (ACC) systems, incorporating adaptive error dynamics to further enhance controller performance and reduce conservativeness. The controller's design utilizes 
Lyapunov functions and Control Barrier Functions (CBFs) to maintain system safety and stability in the presence of potential estimation errors.
The proposed framework guarantees convergence to a fixed conservative bias while maintaining string stability against lead vehicle velocity fluctuations, even under time-varying acceleration.

As demonstrated in Section~\ref{sec:results}, the string stability experiment achieved a gain of 0.57, highlighting the controller's effectiveness in handling these fluctuations even with varying acceleration.
In the CCC experiment, the system maintained an average conservative headway of 0.0317m, while the ACC application exhibited a 0.0364m average conservative headway. These quantitative results emphasize the efficiency of the proposed controller in ensuring safe following distances, in scenarios both with and without V2V communication. The slightly lower headway in the CCC experiment underscores the added benefit of V2V communication, reducing the conservativeness in maintaining safer distances between vehicles.

Moreover, the robustness and safety performance of our solution was further validated through physics engine-based simulations that demonstrated the practical viability of the proposed control framework under realistic environmental conditions. 
In real-world applications, autonomous vehicles are expected to maintain a safe stopping distance of 5.5m. The experiment demonstrated that the vehicles were capable of successfully stopping at distances of 5.548m, 5.5465m, and 5.9357m by the end of the experiment while consistently maintaining a time headway $h > 0$ throughout the experiment.

We identify several promising avenues for future research. These include the incorporation of input constraints, as well as addressing various noise factors such as communication delays, input lags, and packet dropouts in the safety-critical controller design.
The controller proposed in this paper, while focusing on safe and stable cruise control, does not account for the vehicle's physical limitations in practical scenarios. Hence, the integration of input constraints is recognized as a key prospective avenue for future research.
Moreover, exploring the impact of time-varying delays and sensor noise on the estimator design could further enhance its performance and robustness in real-world applications.
Future research directions could also include extending fault tolerance. The current architecture prioritizes safety which will force the platoon to go stale in case any of the preceding vehicles stops. Investigating the potential effects of actuator failures on the system's safety and utilizing the state estimates and consensus to detect faults and adversaries in the platoon can be another key direction for future research.

\bibliographystyle{IEEEtran}
\bibliography{bibtex}

\appendices

\section{}
This appendix section provides essential calculations for the controller and estimator design. Appendix~\ref{sec:A1} presents the necessary conditions for the estimator state matrix to be Hurwitz, and Appendix~\ref{sec:A2} derives one possible solution for the bounds in (\ref{eq:ed}).

\subsection{Estimator Design}\label{sec:A1}
This section outlines the conditions for the estimator gain variables $g_1$, $g_2$, and $g_3$, required to render the estimator state matrix $A$ in (\ref{eq:2}) Hurwitz.
For a $3 \times 3$ matrix to be Hurwitz, the real part of the roots of the characteristic equation must be negative, i.e., $Re(\lambda_i)<0,\, i=1,2,3$. The characteristic equation for the state matrix $A$ is given by:
\begin{equation}
    \lambda^3 -g_1 \lambda^2 - g_2 \lambda - g_3 = (\lambda - \lambda_1)(\lambda - \lambda_2)(\lambda - \lambda_3) = 0
\end{equation}
Upon simplification, the following relations are obtained:
\begin{align}
    g_1 &= \lambda_1 + \lambda_2 + \lambda_3 \nonumber \\
    g_2 &= - (\lambda_1 \lambda_2 + \lambda_1 \lambda_3 + \lambda_2 \lambda_3) \\
    g_3 &= \lambda_1 \lambda_2 \lambda_3 \nonumber
\end{align}
A cubic polynomial can either have three real roots or have one real root and two complex conjugate roots (with $\lambda_2 = \bar \lambda_3$). In both cases, given that $Re(\lambda_i) < 0$, $\lambda_2 + \lambda_3 \in \mathbb R^-$  and $\lambda_2 \lambda_3 \in \mathbb R^+$. Using these properties, the constraints on estimator gain variables are derived as:
\begin{equation}
    g_1,\,g_2,\,g_3 \in \mathbb R^-
\end{equation}
This is a necessary and sufficient condition for the state matrix $A$ to be Hurwitz.

\subsection{Error Dynamics Derivation}\label{sec:A2}
To analyze the bounds on $\dot \epsilon$ in (\ref{eq:ed}), let us define $\gamma$ as:
\begin{equation}
    \gamma (b) = \frac{a (E_u + g_2 \tilde{d}) - \beta(ab)}{a-b} \;\;\text{with} \;\;  b \in \mathbb R^+ - \{a\}
\end{equation}
As the exact value of b is unavailable and only the upper and lower bounds are known, the evolution of $\gamma$ is analyzed against $b$:
\begin{align}
    \frac{\delta \gamma}{\delta b} &= \frac{-a (a-b)\dot \beta(ab) + a (E_u + g_2 \tilde{d}) - \beta(ab)}{(a-b)^2} \nonumber \\
    &= \frac{a (E_u + g_2 \tilde{d} - a \dot \beta(ab)) - \beta(ab) + ab\dot \beta(ab)}{(a-b)^2}
\end{align}

To derive a solution independent of $b$, we choose the class $\mathcal{K}$ function as a linear function, i.e., $\beta(x) = \bar \beta x$. This simplifies the above calculations as follows:
\begin{align}
    \frac{\delta \gamma}{\delta b} = \frac{a (E_u + g_2 \tilde{d} - a \bar \beta)}{(a-b)^2}
\end{align}
This implies that $\gamma$ is constant if $a=0$ or $E_u+g_2\tilde{d} = a\bar\beta$, monotonously increasing if $E_u+g_2\tilde{d}>a\bar\beta$, and monotonously decreasing if $E_u+g_2\tilde{d}<a\bar\beta$.

Therefore, (\ref{eq:ed}) is further simplified as
$\dot \epsilon = \gamma (b) = a \bar \beta = E_u+g_2\tilde{d}$ if $E_u+g_2\tilde{d} = a\bar\beta$, 
and $\dot \epsilon = \gamma(b) = 0$ if $a=0$.
Similarly, for $E_u+g_2\tilde{d}>a\bar\beta$, we obtain:
\begin{align}
    \begin{cases}
        \dot \epsilon = \gamma(b_M) \; ; & a > b \\
        \dot \epsilon = \gamma(a^+) \to - \infty \; ; & a < b \\
        \dot \epsilon = 0 \; ; & a = b
    \end{cases}
\end{align}
where $b_M = \epsilon - v_m$ is the upper bound on $b$ with $v_m$ as the lower bound on $\tilde{v}_1$. Similarly, when $E_u+g_2\tilde{d}<a\bar\beta$, we have $\dot \epsilon = 0$. One possible solution that satisfies these bounds and is independent of $b$ is calculated as in (\ref{eq:21}).

\end{document}